\documentclass[12pt]{article}
\usepackage{latexsym}
\usepackage{epsfig}

\usepackage{graphicx}
\usepackage{epstopdf}

\usepackage{epsfig,amssymb,amsmath,amscd}

\newcommand{\bq}{\begin{equation}}
\newcommand{\eq}{\end{equation}}
\newcommand{\bqa}{\begin{eqnarray}}
\newcommand{\eqa}{\end{eqnarray}}
\newcommand{\ben}{\begin{enumerate}}
\newcommand{\een}{\end{enumerate}}
\newcommand{\bc}{\begin{center}}
\newcommand{\ec}{\end{center}}
\newcommand{\bqb}{\begin{eqnarray*}}
\newcommand{\eqb}{\end{eqnarray*}}

\def\pr#1#2#3{Phys. Rev. ${\bf{#1}}$, #2 (#3)}

\def\pl#1#2#3{Phys. Lett. ${\bf{#1}}$, #2 (#3)}

\def\np#1#2#3{Nucl. Phys. ${\bf{#1}}$, #2 (#3)}

\def\jhep#1#2#3{JHEP ${\bf{#1}}$, #2 (#3)}
\def\epj#1#2#3{Eur. Phys. J. ${\bf{#1}}$, #2 (#3)}

\def\jmp#1#2#3{J. Mod. Phys. ${\bf{#1}}$, #2 (#3)}

\begin{document}
\pagenumbering{arabic}
\thispagestyle{empty}
\def\thefootnote{\fnsymbol{footnote}}
\setcounter{footnote}{1}

\vspace*{2cm}
\begin{flushright}
Feb. 27, 2018\\
 \end{flushright}
\vspace*{1cm}

\begin{center}
{\Large {\bf Polarization effects due to dark matter interaction between
massive standard particles.}}\\
 \vspace{1cm}
{\large F.M. Renard}\\
\vspace{0.2cm}
Laboratoire Univers et Particules de Montpellier,
UMR 5299\\
Universit\'{e} de Montpellier, Place Eug\`{e}ne Bataillon CC072\\
 F-34095 Montpellier Cedex 5, France.\\
\end{center}

\vspace*{1.cm}
\begin{center}
{\bf Abstract}
\end{center}

We propose deeper tests of the existence of DM interactions between heavy particles
in $e^+e^-\to t\bar t, ZZ, W^+W^-$ by looking at the effects on final state polarization. We show that indeed $t$, $W$ and $Z$ polarization are particularly
sensitive to the structure of these interactions, to their relation with the origin
of the masses and to the quantum numbers of the possibly exchanged dark particles.

\vspace{0.5cm}

\def\thefootnote{\arabic{footnote}}
\setcounter{footnote}{0}
\clearpage

\section{INTRODUCTION}

In previous papers \cite{DMmass, DMexch} we have assumed that
massive SM particles ($t,Z,W,...$) are especially connected to dark matter (DM)
whose properties
\footnote{We thank Mike Cavedon for interesting remarks on this subject.}
have been recently reviewed in \cite{rev1}.\\
We want now to explore more deeply what modifications this may generate
in the production mechanism of heavy particles through final state interactions.\\
We reconsider the simplest processes $e^+e^-\to t\bar t, ZZ, W^+W^-$ already treated
in \cite{DMexch}.\\
In the standard model (SM) these three processes have remarkable specific polarization properties at high energy. We  want to see how the polarizations of $t,Z,W,...$ can reflect the basic features of a final DM interaction.

The simplest example is $e^+e^-\to t\bar t$ with the final exchange (in the t channel) of a single particle. Neglecting ${m^2_t\over s}$ terms, the SM production
amplitudes satisfy $\tau=-\tau'$ (the $t$ and $\bar t$ helicities). Then, a final DM interaction
with vector t-channel exchange would ensure helicity conservation (HC) and keep the ordering of the magnitudes of
the $\tau=\pm{1\over2}$ amplitudes, whereas scalar exchange with an helicity violation (HV) effect would interchange these magnitudes and produce a totally different top quark polarization.
The actual DM final state interaction may be more complex and involve multiparticle exchanges.
In any case our illustrations will suggest that a measurement of the top quark polarization should be instructive about the DM dynamics.\\

The situation is totally different in $e^+e^-\to W^+W^-$ and in $e^+e^-\to ZZ$.
The longitudinal components whose existence is directly connected with
the origin of the masses should be particularly affected by DM interactions
if DM is implied in this connection.\\
In SM the well-known feature is the gauge cancellation of the amplitudes
for longitudinal $W^{\pm}_L$ or $Z_L$ components at high energy (neglecting
${m^2_{W,Z}\over s}$ terms).
In $e^+e^-\to W^+W^-$  the TT part is larger than the LL one, itself larger than TL.
In $e^+e^-\to ZZ$,  TT  is larger than TL,  itself larger than LL.\\
These properties (in particular these orderings) would be immediately affected, even by minor modifications of the SM contributions, when final DM interactions would not be identical for all helicities.\\
We will illustrate these possibilities with arbitrary examples leading to modifications of the final polarizations.\\

Our conclusion is that detailed analyses of these processes may allow to identify the involved dynamical features of the DM interactions (or of their portals, for definition
see  \cite{Portal}) and possibly distinguish them from other types of
SM modifications.\\

Contents:  Sect.2  is devoted to $e^+e^-\to t\bar t$ and  Sect.3  to
$e^+e^-\to W^+W^-$ and $e^+e^-\to ZZ$. A summary of the results, of
their implications and possible future developments are given in Sect.4.

\section{Top quark polarization}

At Born level the $t\bar t$ pair is produced by $e^+e^-$ annihilation
through photon and $Z$ whose properties are well-known, see for example
\cite{bookee}.
Our aim is to see how the various polarization observables
would be affected by final state DM interactions.
We will illustrate the energy and angular dependences of the top
quark polarization
\bq
P_t(s,\theta)={\sigma(\tau=+{1\over2})-\sigma(\tau=-{1\over2})\over
\sigma(\tau=+{1\over2})+\sigma(\tau=-{1\over2})}
~\label{pt}~~\eq
for unpolarized initial beams,
and of $P^L_t$ for left-handed polarized $e^-$  beams,
$P^R_t$ for right-handed polarized $e^-$  beams,
and of the corresponding polarization asymmetry $A_{LR}$ .\\
The basic SM predictions can be seen in each of the Figs.3-10.
They result from the peculiar combinations of the photon and $Z$
couplings to $e^+e^-$ and to $t\bar t$, see \cite{bookee}.\\
At high energy, for $s>m^2_Z,m^2_t$,
the amplitudes are controlled by the coefficients
\bq
G(e,t)=q_eq_t+g^{L,R}_{Ze}g_{Zt^{+,-}}
\eq
with $g^{L}_{Ze}={-1+2s^2_w\over2s_wc_w}$, $g^{R}_{Ze}={s_W\over c_W}$,
$g_{Zt^{+}}=-{2s_w\over3c_w}$, $g_{Zt^{-}}={1-{4s^2_w\over3}\over2s_wc_w}$.
In SM one has approximatively
$G(e_Lt_-)=G(e_Rt_+)$ and $G(e_Rt_-)=G(e_Lt_+)$,  
(this would be exact for $s^2_W={1\over4}$).\\
This explains the opposite values obtained for $P^{L}_t$ and $P^{R}_t$.\\
Note that for unpolarized $e^+e^-$ beams the resulting $P_t$ magnitude is much smaller
(it would vanish for $s^2_W={1\over4}$).\\
The angular dependences arises from the final top, antitop spinors
coeficients $\sin{\theta\over2}$ and $\cos{\theta\over2}$.\\

We want now to see how these properties would be modified by final state
DM interaction accordingly with the pictures
of Fig.1 as in \cite{DMexch}.\\

We write
\bq
\sigma(\tau,\tau')=\sum_{\lambda,\lambda'}\sigma^{SM}(\lambda,\lambda')
C(\lambda,\lambda',\tau,\tau')
\eq
where the DM effect (the black box in Fig.1) is represented by the function $C$ which should in general
depend on the initial and final helicities.\\
We want first to see how one could differentiate the HC from the HV
DM interactions and we will then use more specific models.\\

In a first step we globally affect the Born helicity amplitudes
by a correction factor $C(\lambda,\lambda',\tau,\tau'$ ) which satisfies either helicity conservation
$(\lambda=\tau ; \lambda'=\tau')$   or  helicity violation
$(\lambda=-\tau ; \lambda'=-\tau')$.\\
As an example, for illustration we use in each of these cases

\bq
C={m^2_{t}\over m^2_0}~ln{-s\over 4m^2_{t}} ~~, \label{Fs}
\eq
\noindent
(like in \cite{DMexch} and  \cite{CSMrev}) with $m_0=0.5$ TeV.\\

The resulting changes in the various polarization observables can be seen in
Figs.3-6.\\
As expected in the HC case the resulting top polarization is similar to the SM
one. This arises from the fact that all the amplitudes with different top
helicities are modified with a common factor which disappears in the ratio,
eq.(\ref{pt}), defining $P_t$.\\
On the opposite the HV case, which interchanges the two top helicities,
leads to a sign change of $P_t$ in all cases illustrated in Figs.3-6.\\

In a second step we modelize DM effects
by considering separately Vector or Scalar DM exchanges
according to diagrams of Fig.2.
At high energy it is well known
that vector(scalar) exchange in the t-channel will produce pure HC(HV) effects .
But additional contributions from the s-channel arise for specific helicities
$(\lambda=-\lambda' ; \tau=-\tau')$ for the vector case and
$(\lambda=\lambda' ; \tau=\tau')$ for the scalar case, which will perturb the above
t-channel properties. The resulting effects on the top quark polarization are
shown in Figs.7-10.\\
The total vector effects although somewhat numerically different have
similarities with the SM case. On the opposite the total scalar case generates
very different energy and angular dependences due to the addition of the
t-channel and s-channel properties.\\

These illustrations show that, indeed, measurements of top quark polarization
should allow to get important informations about a possible final state interaction.
The DM exchanges occuring in Fig.2 maybe more complex (multiparticle)
than the above simple vector and scalar cases. It may also arise through
simple portals and in that case may behave like in our illustrations.

In any case the corresponding analysis
may require a maximal number of $e^+e^-\to t\bar t$ observables, but clearly
the top quark polarization should play an important role.\\

Precise constraints on weak DM effects from experimental measurements
will require to take into account the high order SM corrections.\\

For comparison it will be interesting to look at other types of non
standard effects for example those due to top quark (and Higgs boson) compositeness (see ref.\cite{comp},
\cite{Hcomp2},\cite{Hcomp3},\cite{Hcomp4},\cite{partialcomp}, \cite{Tait},
\cite{trcomp})
and it will be important to find specific signals allowing to distinguish
this origin from the other ones.\\

Staying at the level of effective description the experimental results
may also be analyzed in terms
of specific anomalous top couplings, see for example \cite{anomtt1, anomtt2}.\\

Reviews of the experimental possibilities for $e^+e^-\to t\bar t$
can be found in \cite{Moortgat}, \cite{Craig}.\\

\section{$W$ and $Z$ polarization}

\subsection{$e^+e^-\to W^+W^-$}

In the same spirit we now consider the process $e^+e^-\to W^+W^-$,
the DM final state interaction being
described with the corresponding diagrams of Figs.1 and 2.\\
The phenomenology is however different from the one of the $t\bar t$ case.\\
The special point is now the behaviour of the longitudinal $W$ components.
It is well-known that without the specific SM gauge cancellation they would
strongly increase (like ${E\over m}$). So any, even small, modification of
the contributions could perturb these cancellations and lead to an increase
of the $W_L$ amplitudes.\\
In Figs.11,12 we show the energy and angular depencences of the  $W^+_TW^-_T$,
$W^+_TW^-_L$, $W^+_LW^-_L$ production cross sections, with solid lines for the SM
case.\\
As a simple example we then add a DM final state contribution with (t- and s- channel) exchanges of a scalar contribution  as described in Fig.2.
Indeed it immediately leads
to an increase of the $W^+_LW^-_L$ cross section at high energy and also generates
a strong modification of the corresponding angular distribution.\\
In general only smaller
modifications appear for the $W^+_TW^-_L$ and the $W^+_TW^-_T$ cross sections.\\

The possibilities of non standard effects at the future ILC have been reviewed in \cite{WWILC}.
Here also the comparison with effects of anomalous couplings may be
done, see for ex.\cite{WWanom1, WWanom2}.

\subsection{$e^+e^-\to ZZ$}

We do a similar treatment of this process acording to Figs.1,2.\\
The situation is similar to the one of $e^+e^-\to W^+W^-$ except that
the SM Born contribution is even much simpler with only the t- and u-channel
electron exchange amplitudes which much largely cancel at high energy
in the case of longitudinal $Z_L$ amplitudes. The resulting TL and LL SM cross sections indeed quickly decrease with the energy.\\
Consequently a small additional contribution,
although invisible in the TT part, immediately strongly modifies the LL
part as can be seen in Figs.11,12 (and at a weaker level also the TL one).\\

More detailed analyses should be done with various types of DM exchanges
between $WW$ and between $ZZ$ states in order to see how polarization measurements
could caracterize the DM properties.\\

\section{CONCLUSION}

We have more deeply explored the possible tests of the existence of a final
$t\bar t, ZZ, W^+W^-$ interaction between heavy particles due to their connection with
DM. We have shown that strong modifications of the $t,W,Z$ polarizations may
appear and reveal the structure of the DM interactions. Simple components with HC or HV amplitudes immediately lead to specific $P_t$ modifications that we have illustrated in the $e^{\pm}$ unpolarized, $e^-_L$, $e^-_R$ cases.
Such components may be generated
by Vector or Scalar t-channel exchanges (for examples portals to DM). A more complex combination of multiparticle
exchanges may lead to other effects, but what we wanted to show is that $P_t$
measurements should be useful for analyzing the structure of this final interaction.\\
The situation of the $W^{\pm}$, $Z$ polarizations is totally different because of the very strong sensitivity of the  $W^{\pm}_L$, $Z_L$ amplitudes to any modification
of the SM ones which satisfy the typical gauge cancellations at high energy. We
have illustrated how TL, LL cross sections will be modified (whereas the TT one is not)
by final DM interactions with the t-channel, s-channel exchanges examples already used in the top quark case.\\

Future developments may consist in searching a way to distinguish these DM effects from other BSM ones, like compositeness and in general effective anomalous couplings.\\

It seems clear that polarization measurezments should certainly constitute an important tool for that purpose.\\

We have devoted the present study to the simplest $e^+e^-$ processes.

Other processes may also be studied in
photon-photon collisions \cite{gammagamma}, and in hadronic processes
\cite{Contino}, \cite{Richard}.

\newpage
\begin{figure}[p]
\vspace{-0cm}
\[
\hspace{-2cm}\epsfig{file=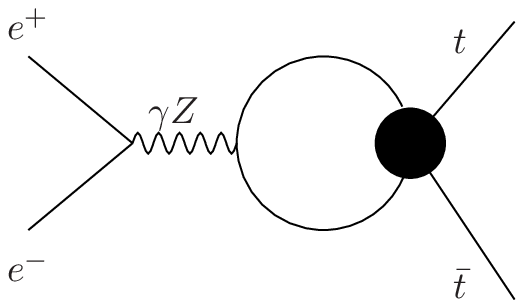 , height=30.cm}
\]\\
\vspace{-28cm}
\[
\hspace{-2cm}\epsfig{file=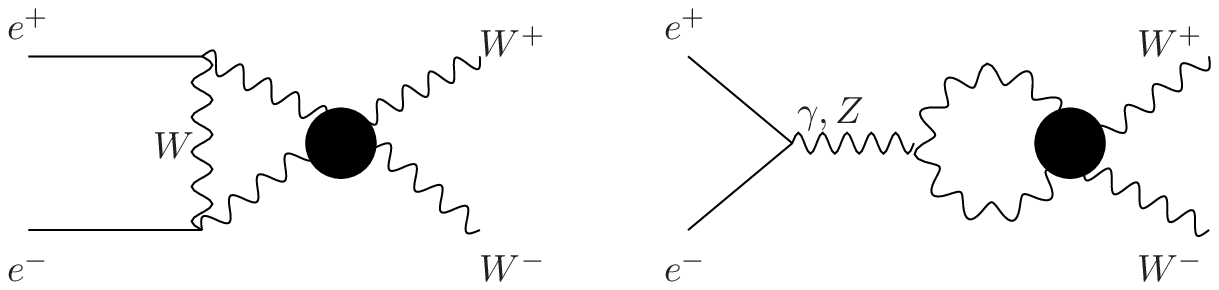 , height=30.cm}
\]\\
\vspace{-28cm}
\[
\hspace{-2cm}\epsfig{file=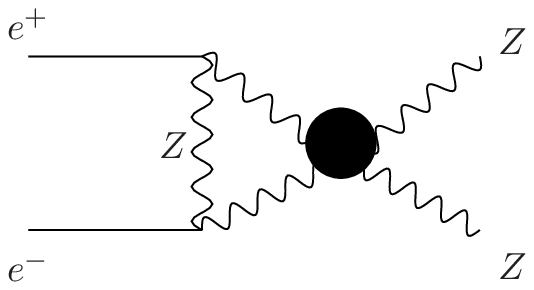 , height=30.cm}
\]\\
\vspace{-20cm}
\caption[1] {Simple processes with production of a pair of
massive particles submitted to final state DM interaction; $ZZ$ symmetrization is applied.}
\end{figure}

\clearpage

\begin{figure}[p]
\vspace{-10cm}
\[
\hspace{-2cm}\epsfig{file=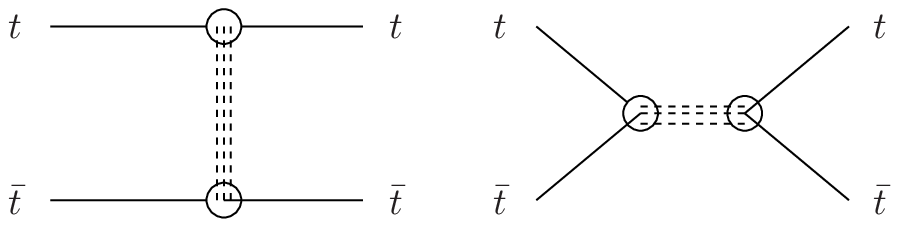 , height=30.cm}
\]\\
\vspace{-28cm}
\[
\hspace{-2cm}\epsfig{file=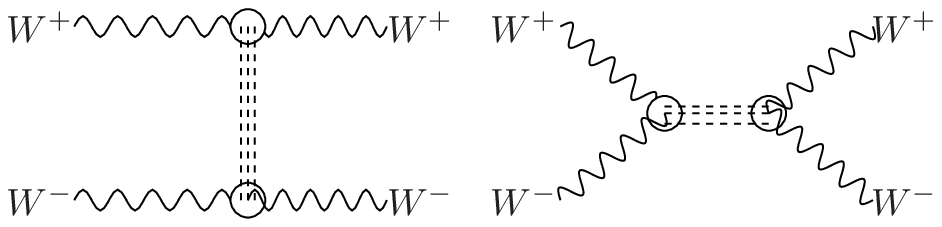 , height=30.cm}
\]\\
\vspace{-28cm}
\[
\hspace{-2cm}\epsfig{file=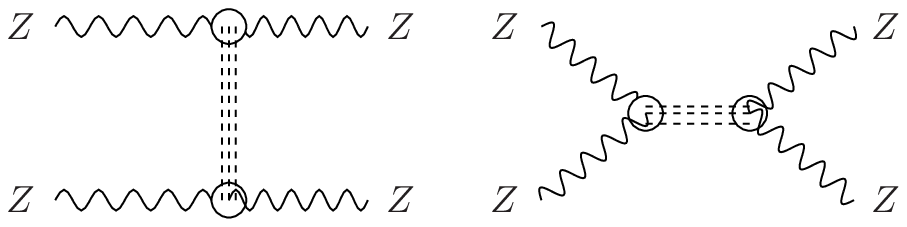 , height=30.cm}
\]\\
\vspace{-15cm}
\caption[1] {Examples of final state DM interaction with s,t,u channel
exchanges; $ZZ$ symmetrization is applied.}
\end{figure}

\clearpage

\begin{figure}[p]
\vspace{-0cm}
\[
\hspace{-2cm}\epsfig{file=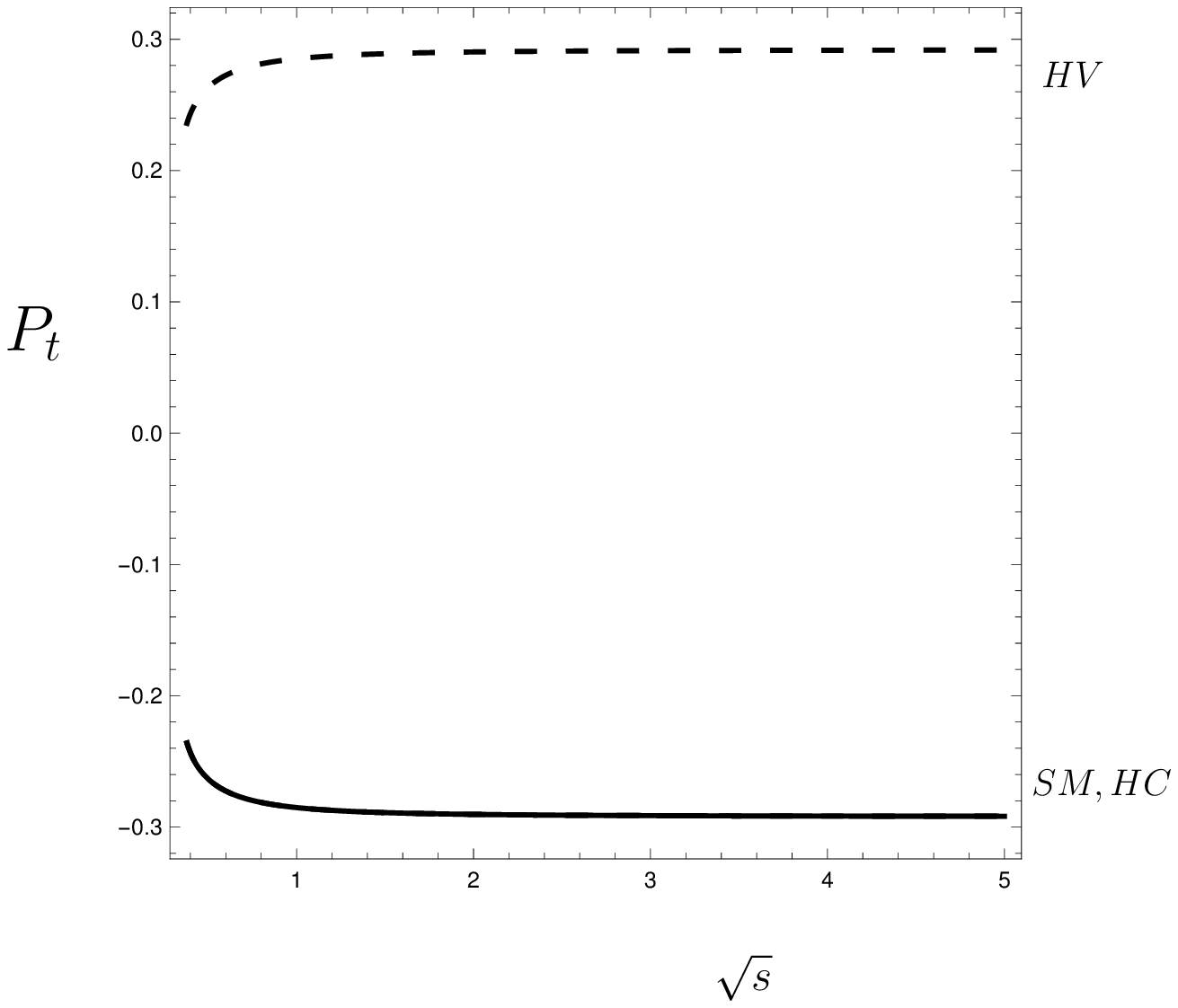 , height=20.cm}
\]\\
\vspace{-13cm}
\[
\hspace{-2cm}\epsfig{file=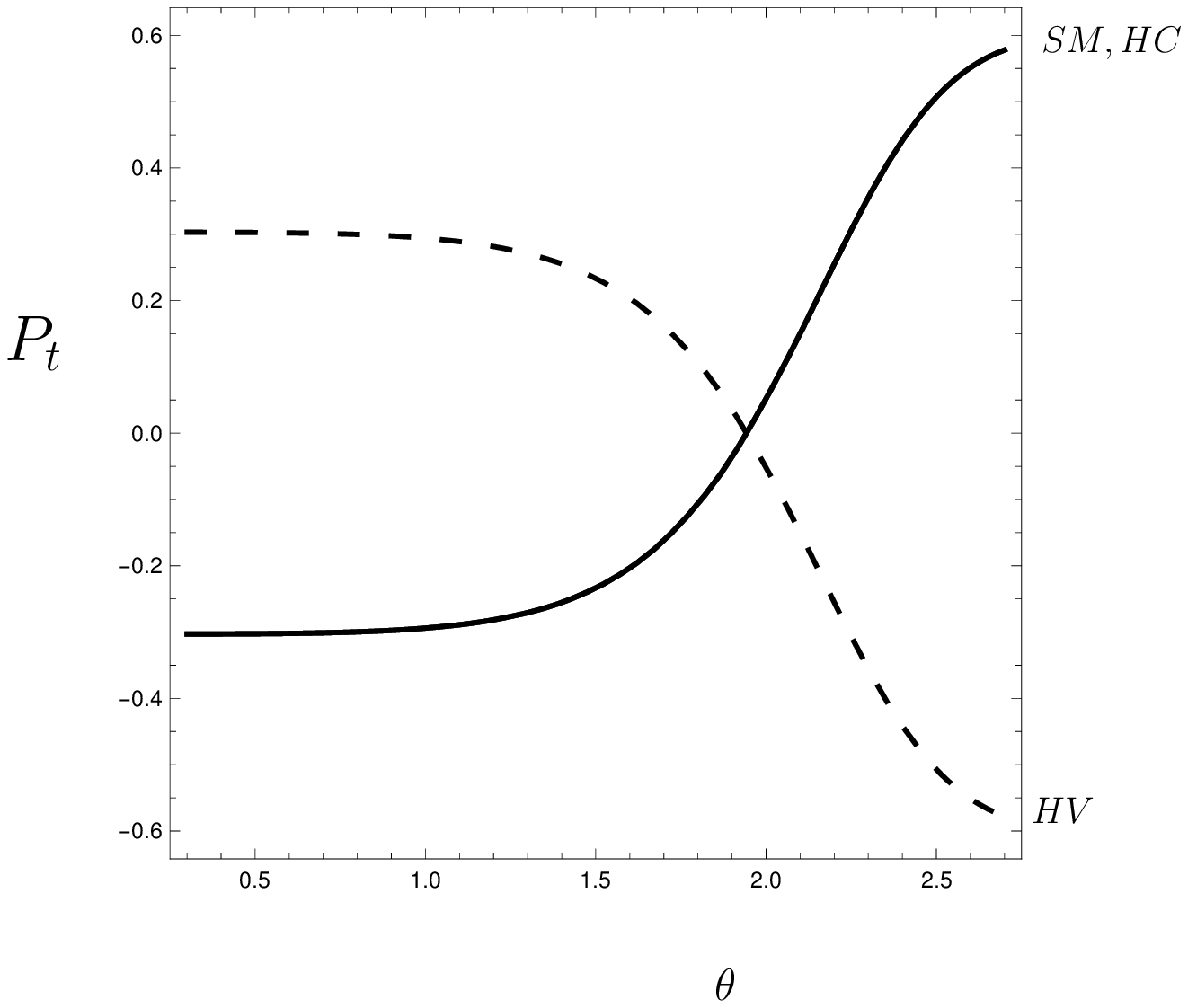 , height=20.cm}
\]\\
\vspace{-10cm}
\caption[1] {Energy and angular (at $\sqrt{s}=5$ TeV) dependences of the top quark polarization
in the $e^{\pm}$ unpolarized case,
in SM,  with HC dark matter and with HV dark matter.}
\end{figure}

\clearpage

\begin{figure}[p]
\vspace{-0cm}
\[
\hspace{-2cm}\epsfig{file=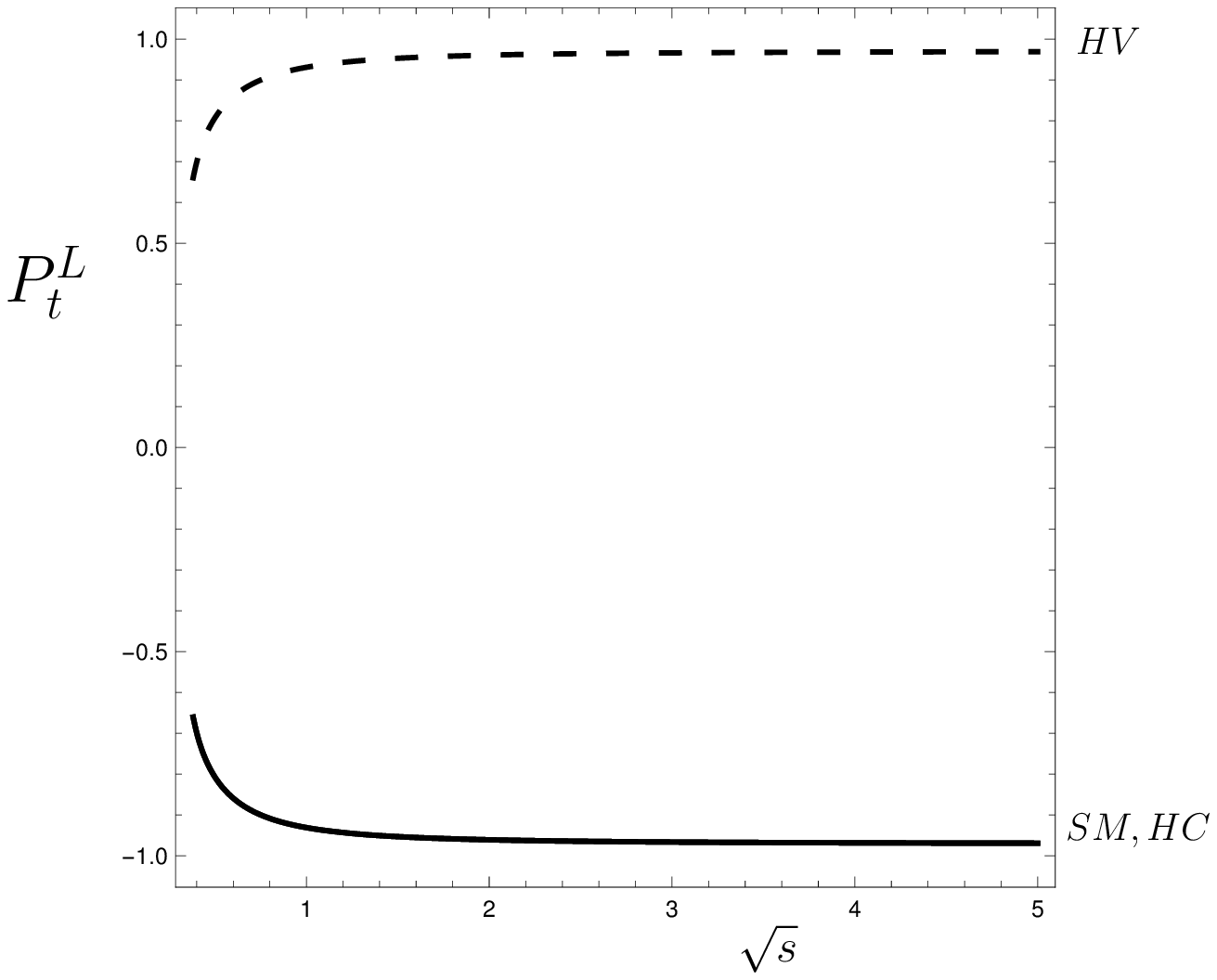 , height=20.cm}
\]\\
\vspace{-13cm}
\[
\hspace{-2cm}\epsfig{file=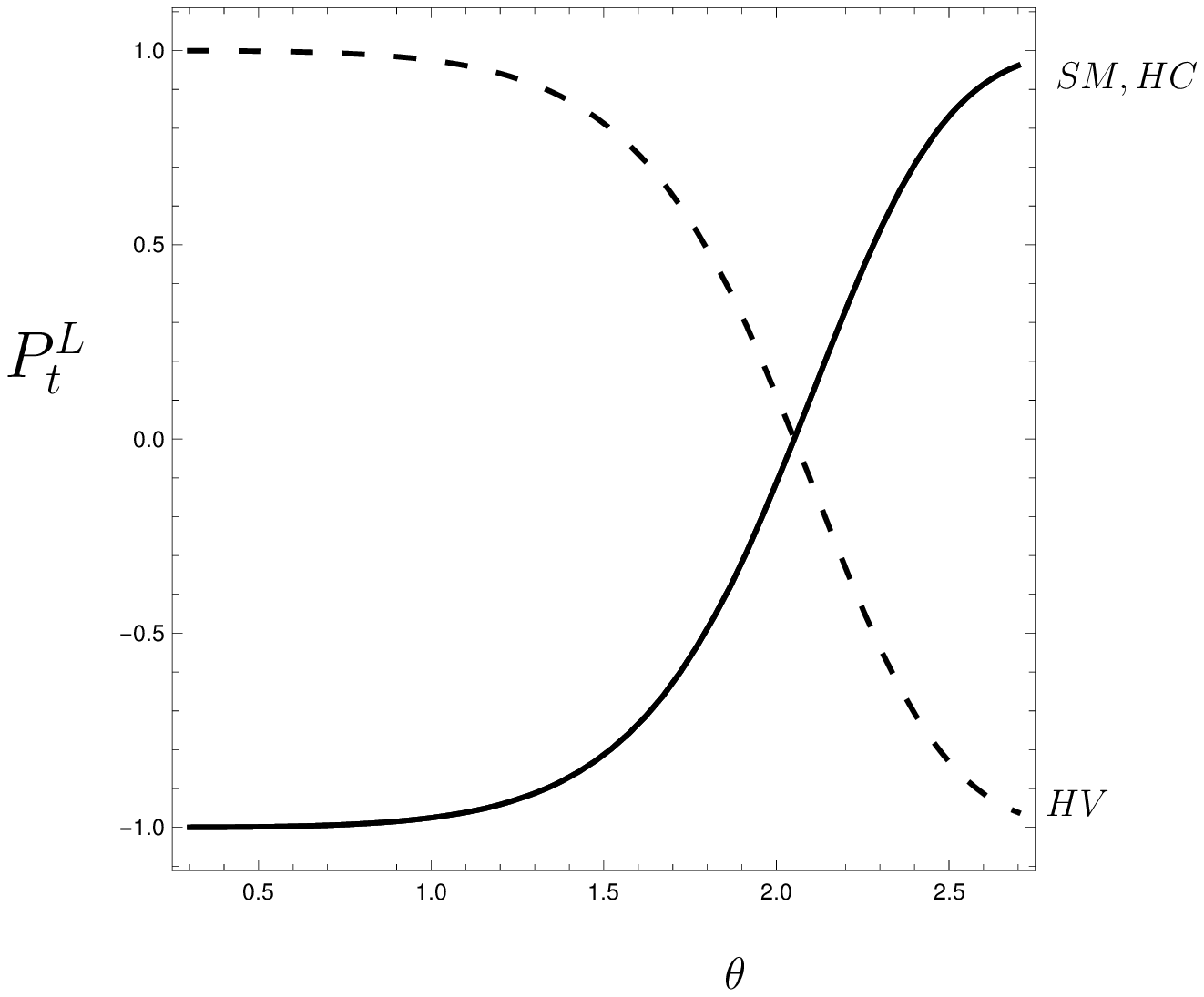 , height=20.cm}
\]\\
\vspace{-10cm}
\caption[1] {Energy and angular (at $\sqrt{s}=5$ TeV) dependences of the top quark polarization
in the $e^{-}_L$ polarized case,
in SM,  with HC dark matter and with HV dark matter.}
\end{figure}
\clearpage

\begin{figure}[p]
\vspace{-0cm}
\[
\hspace{-2cm}\epsfig{file=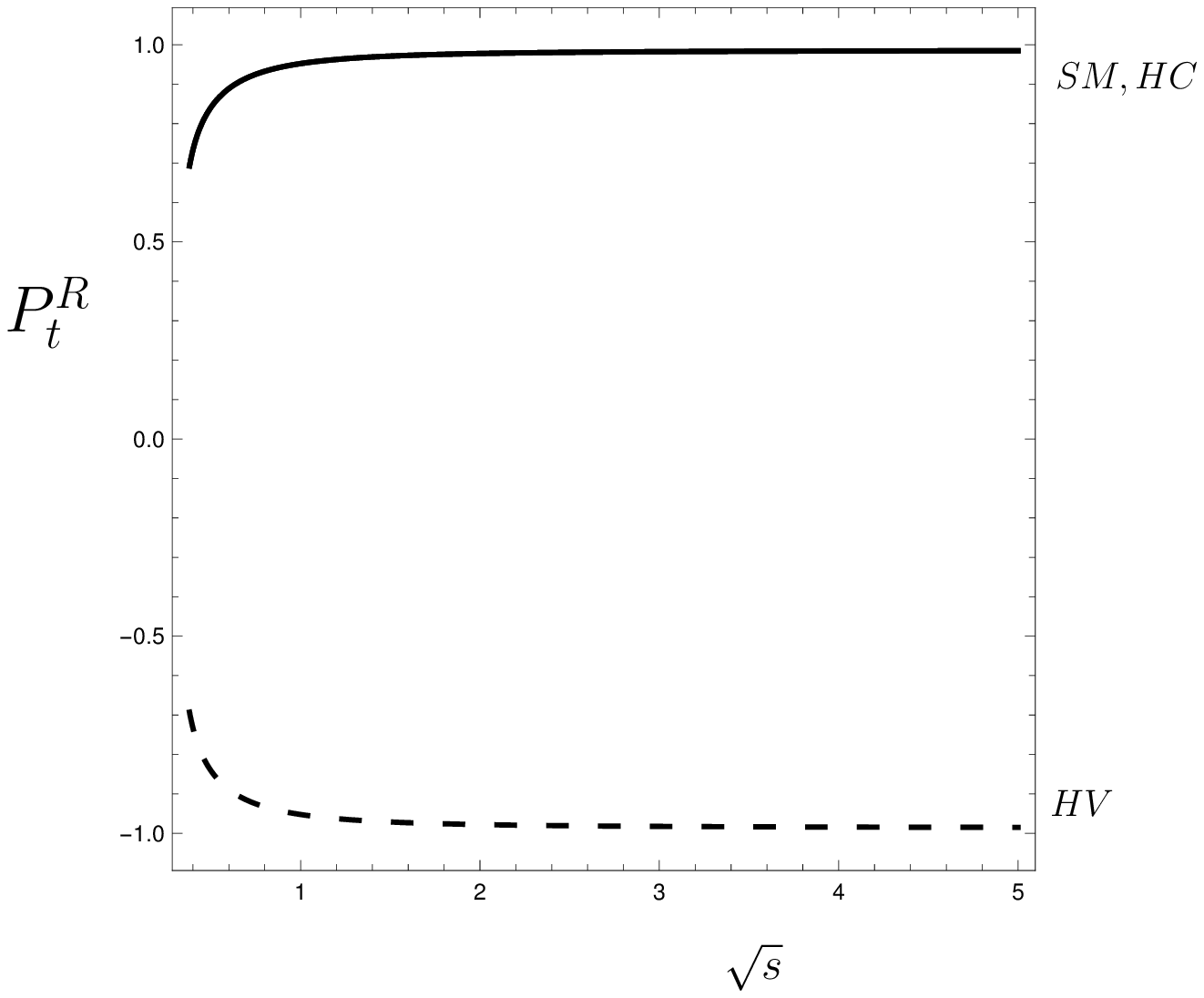 , height=20.cm}
\]\\
\vspace{-13cm}
\[
\hspace{-2cm}\epsfig{file=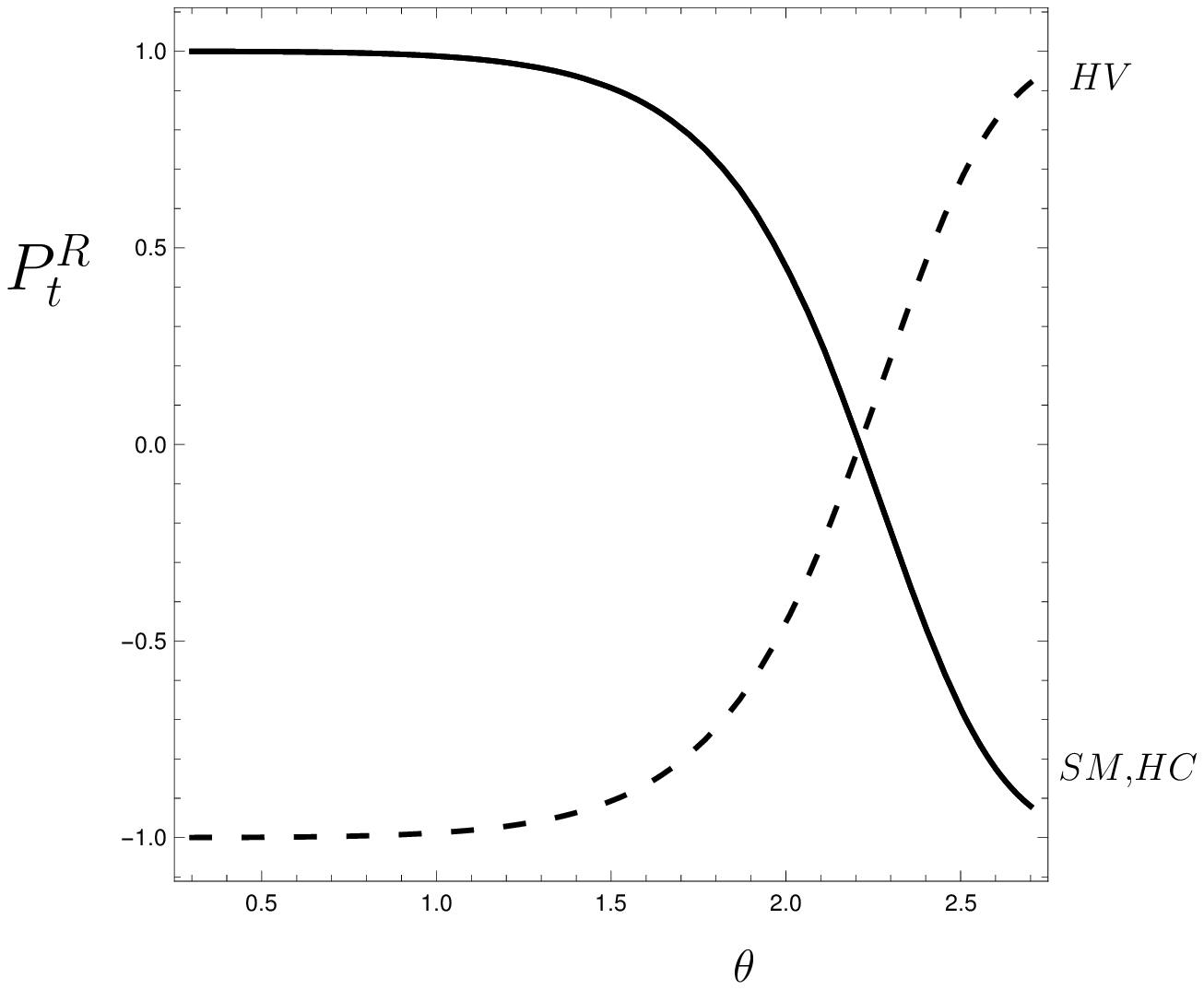 , height=20.cm}
\]\\
\vspace{-10cm}
\caption[1] {Energy and angular (at $\sqrt{s}=5$ TeV) dependences of the top quark polarization
in the $e^{-}_R$ polarized case,
in SM,  with HC dark matter and with HV dark matter.}
\end{figure}
\clearpage

\begin{figure}[p]
\vspace{-0cm}
\[
\hspace{-2cm}\epsfig{file=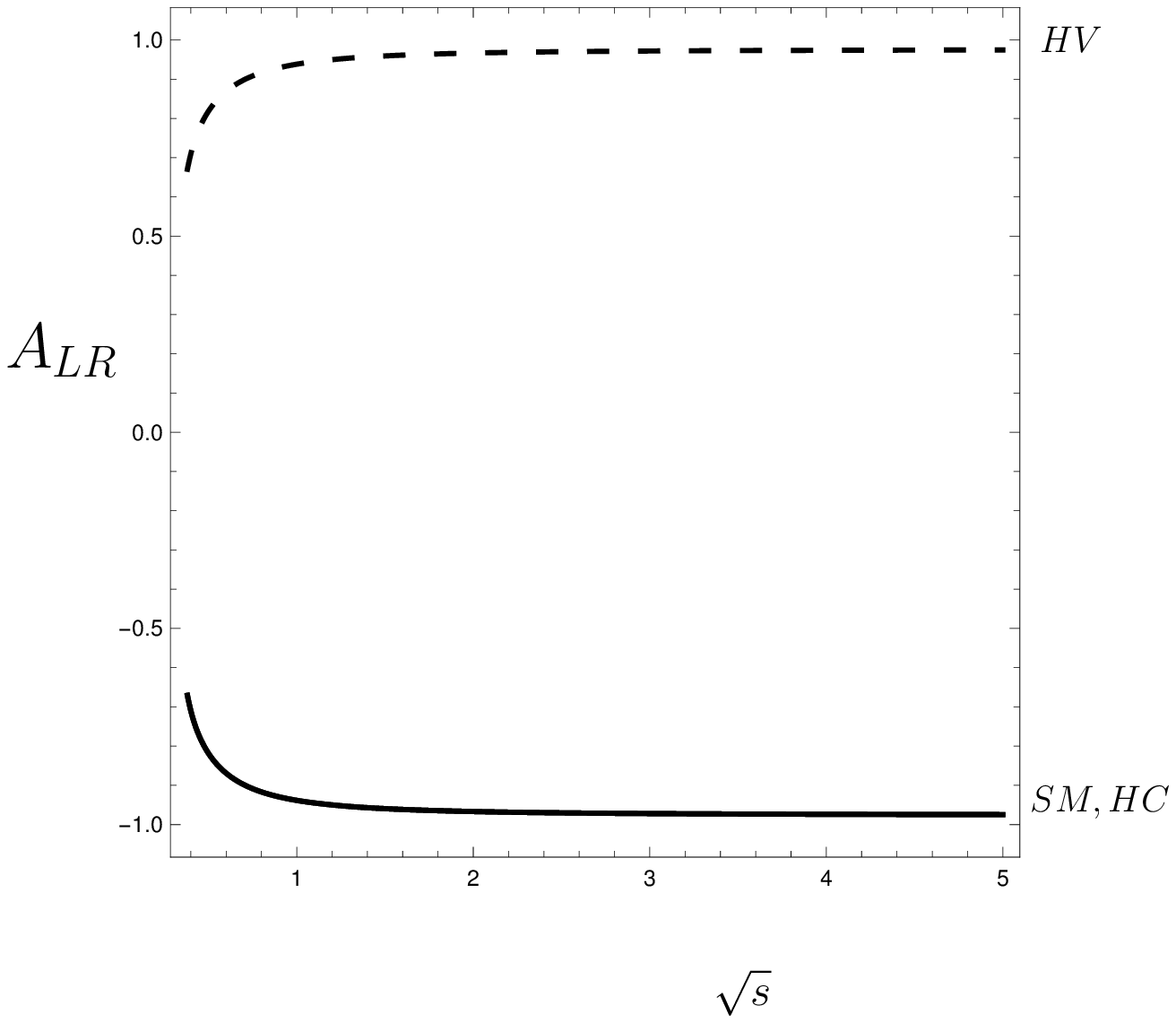 , height=20.cm}
\]\\
\vspace{-13cm}
\[
\hspace{-2cm}\epsfig{file=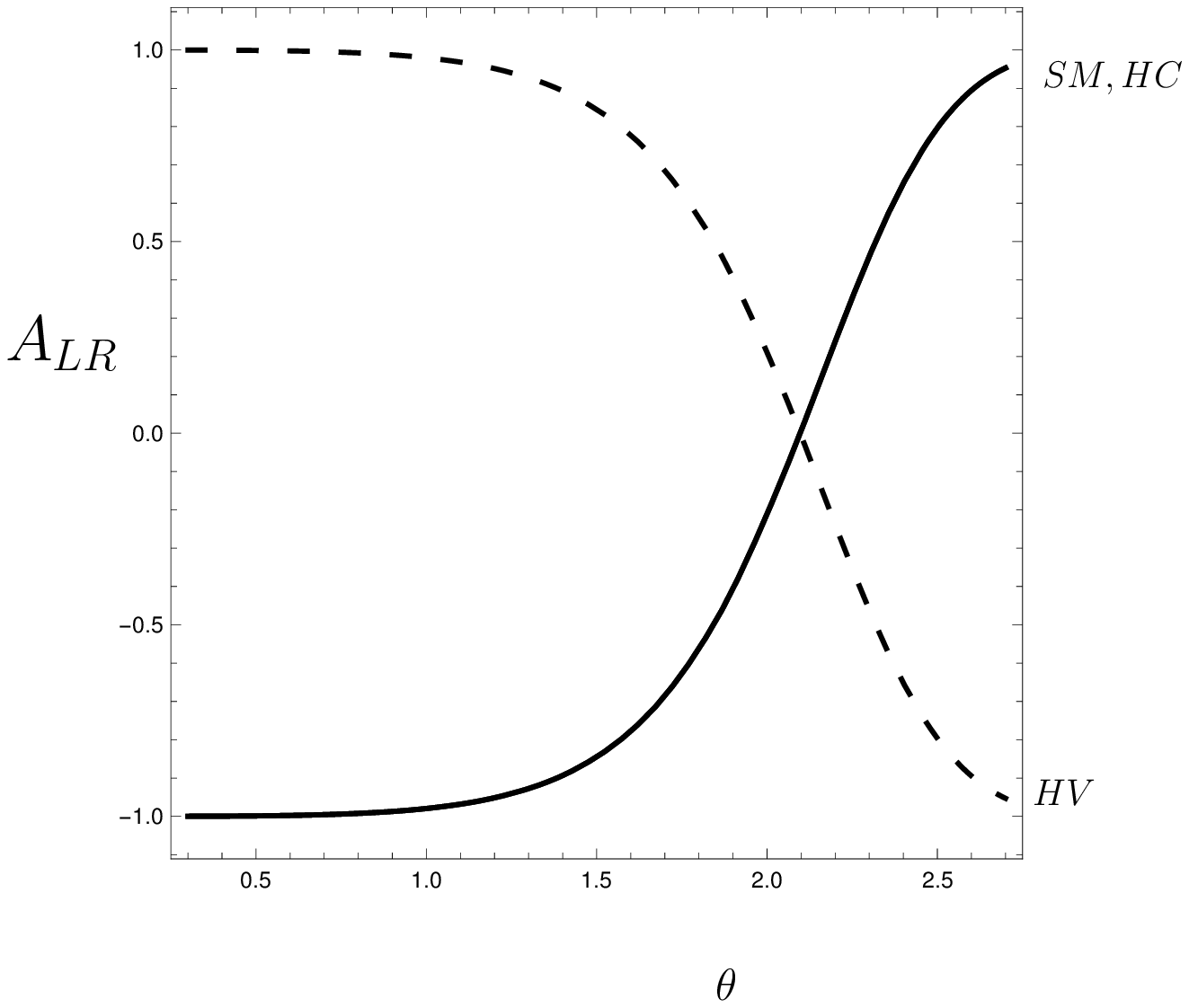 , height=20.cm}
\]\\
\vspace{-10cm}
\caption[1] {Energy and angular (at $\sqrt{s}=5$ TeV) dependences of the $e^{-}_L$-$e^{-}_R$
asymmetry of the top quark polarization,
in SM,  with HC dark matter and with HV dark matter.}
\end{figure}
\clearpage

\begin{figure}[p]
\vspace{-0cm}
\[
\hspace{-2cm}\epsfig{file=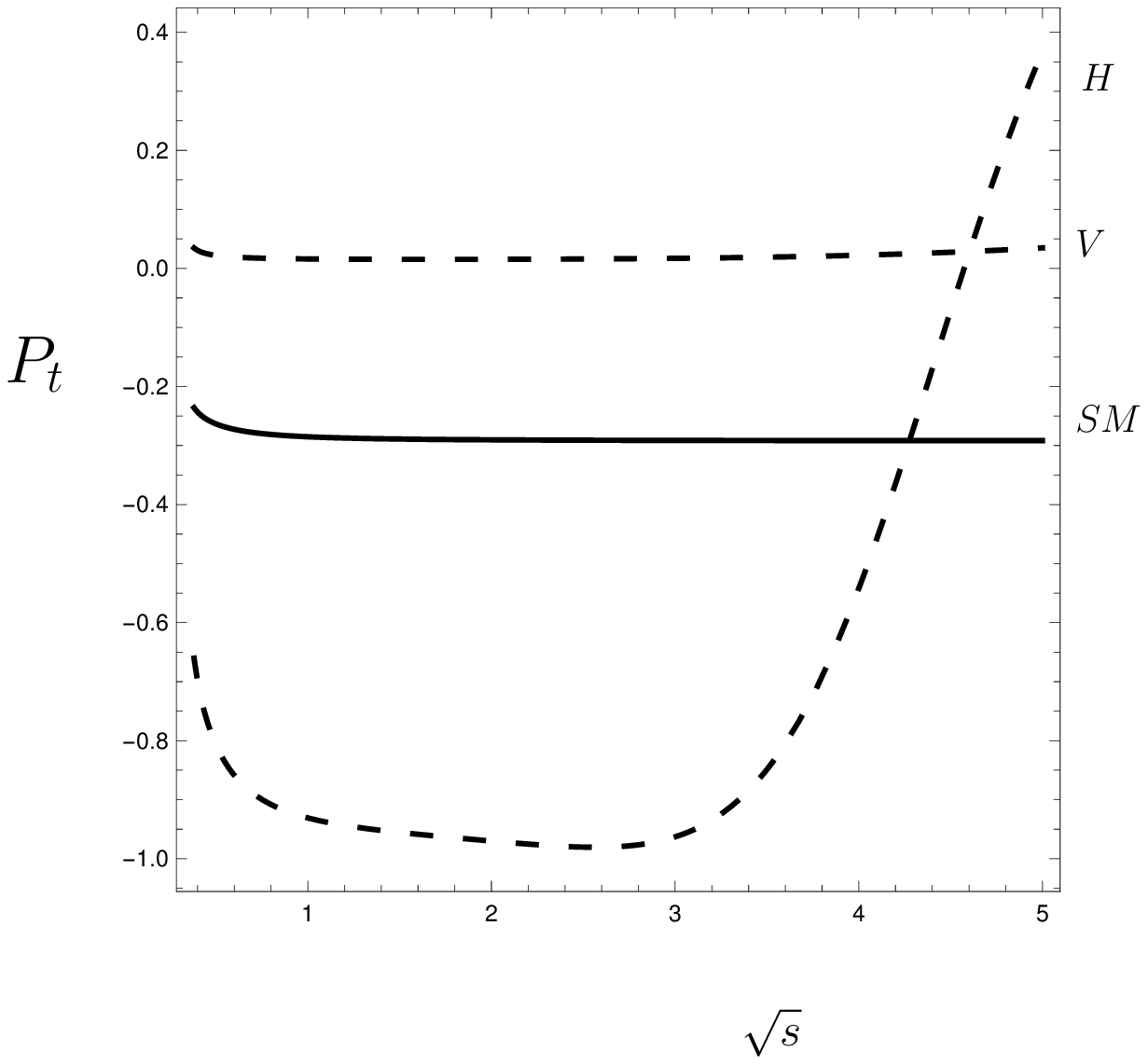 , height=20.cm}
\]\\
\vspace{-13cm}
\[
\hspace{-2cm}\epsfig{file=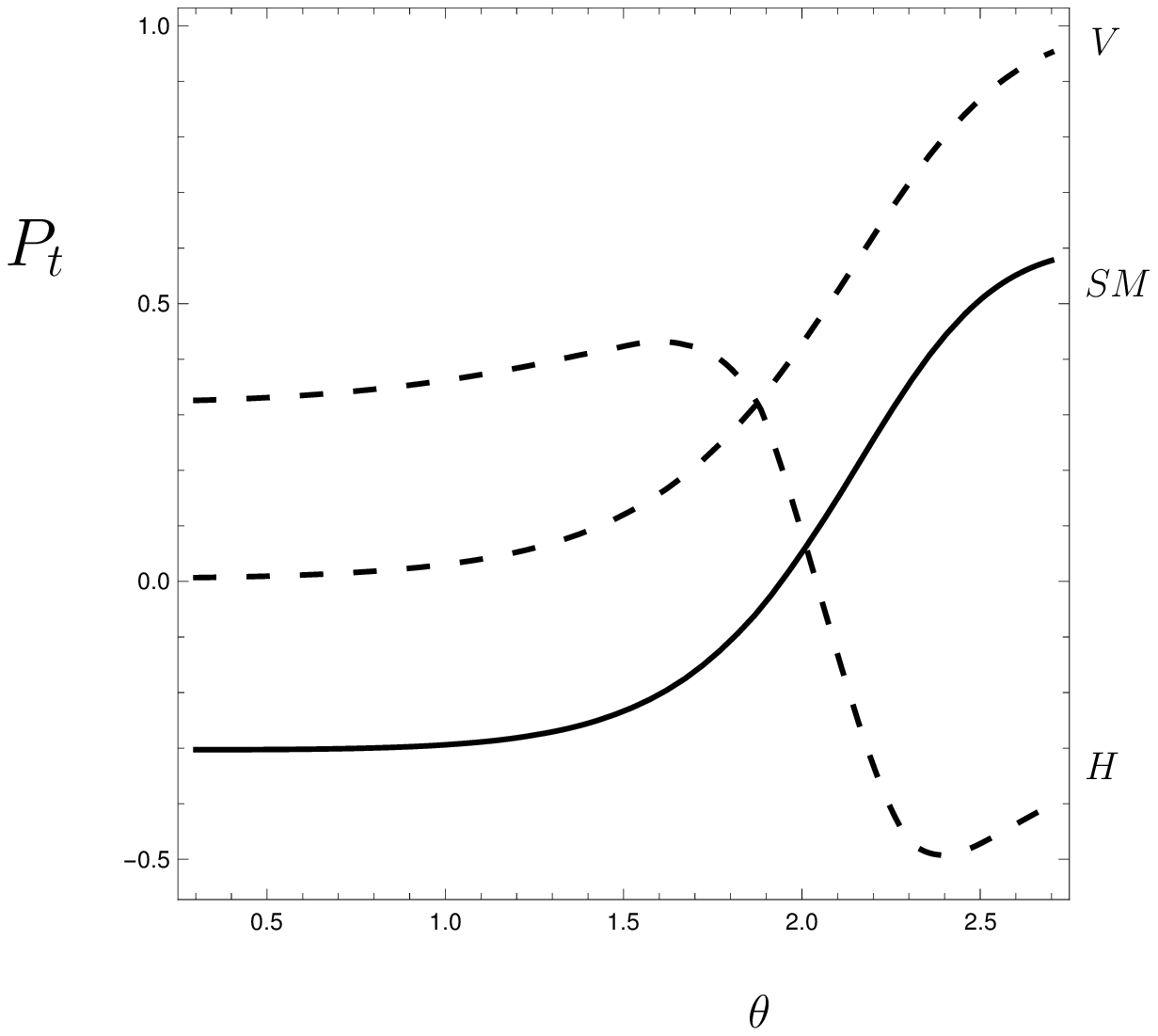 , height=20.cm}
\]\\
\vspace{-10cm}
\caption[1] {Energy and angular (at $\sqrt{s}=5$ TeV) dependences of the top quark polarization
in the $e^{\pm}$ unpolarized case,
in SM,  (H) with scalar Higgs type dark matter and (V) with vector dark matter.}
\end{figure}
\clearpage

\begin{figure}[p]
\vspace{-0cm}
\[
\hspace{-2cm}\epsfig{file=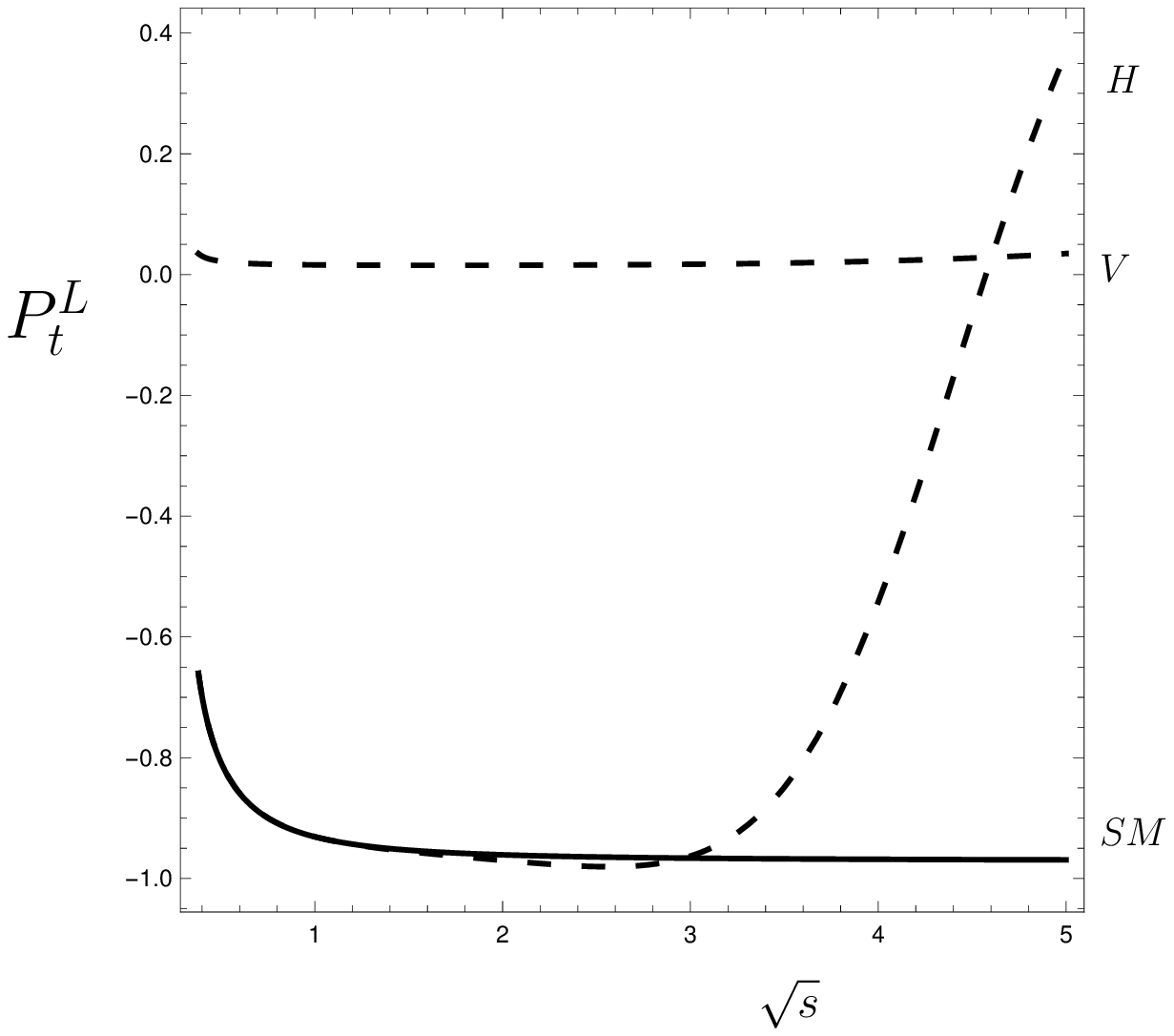 , height=20.cm}
\]\\
\vspace{-13cm}
\[
\hspace{-2cm}\epsfig{file=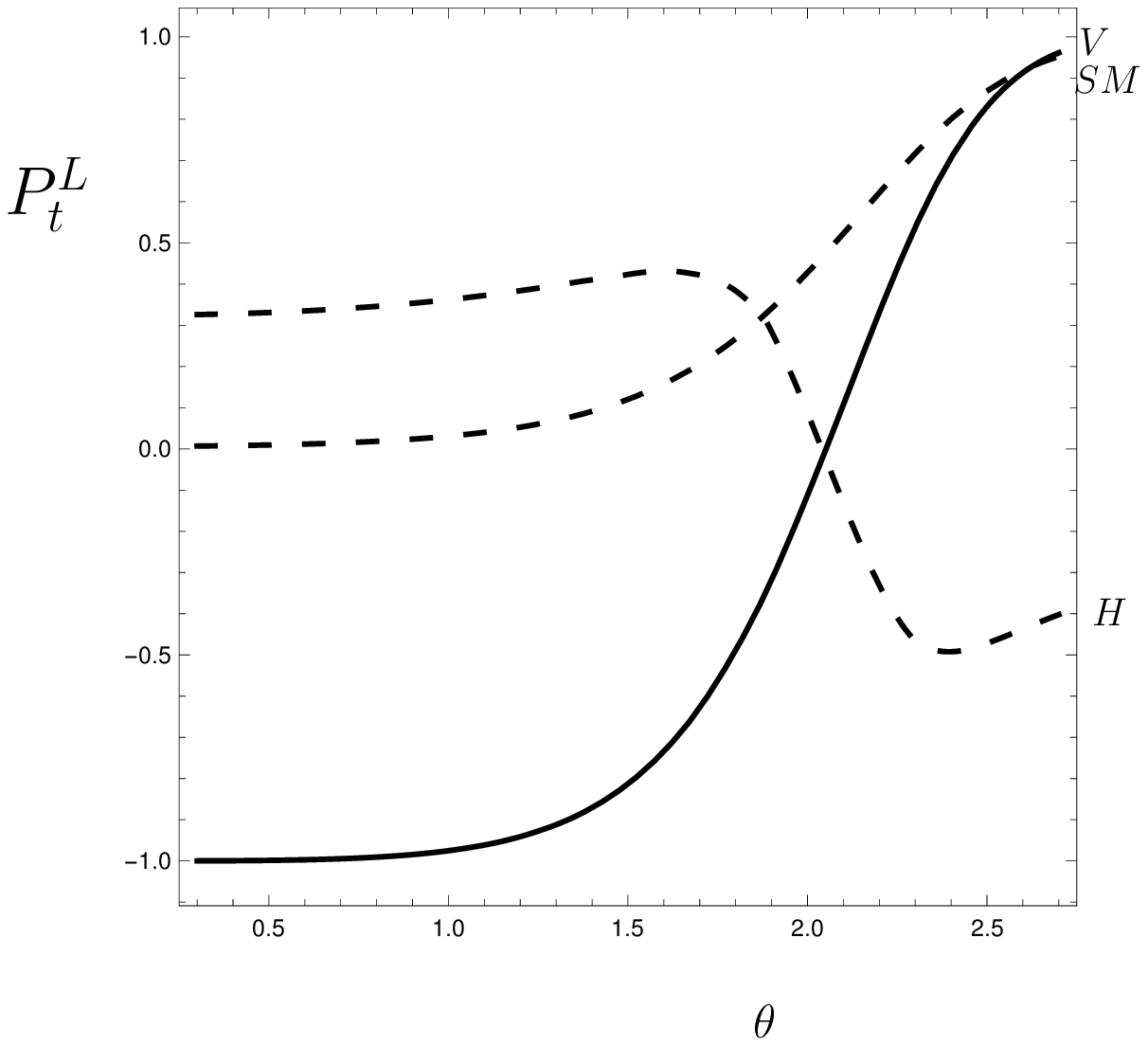 , height=20.cm}
\]\\
\vspace{-10cm}
\caption[1] {Energy and angular (at $\sqrt{s}=5$ TeV) dependences of the top quark polarization
in the $e^{-}_L$ polarized case,
in SM,  (H) with scalar Higgs type dark matter and (V) with vector dark matter.}
\end{figure}
\clearpage

\begin{figure}[p]
\vspace{-0cm}
\[
\hspace{-2cm}\epsfig{file=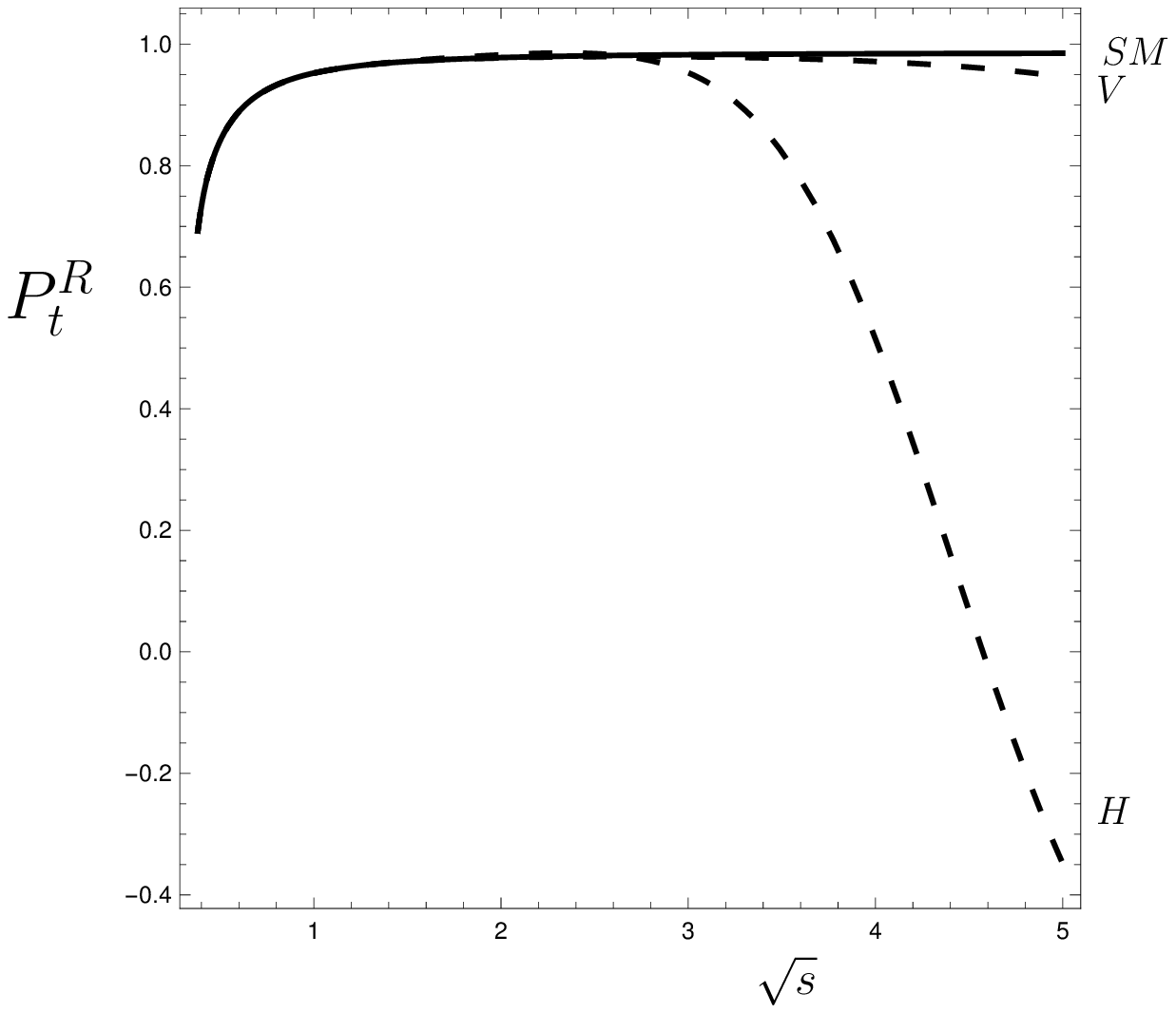 , height=20.cm}
\]\\
\vspace{-13cm}
\[
\hspace{-2cm}\epsfig{file=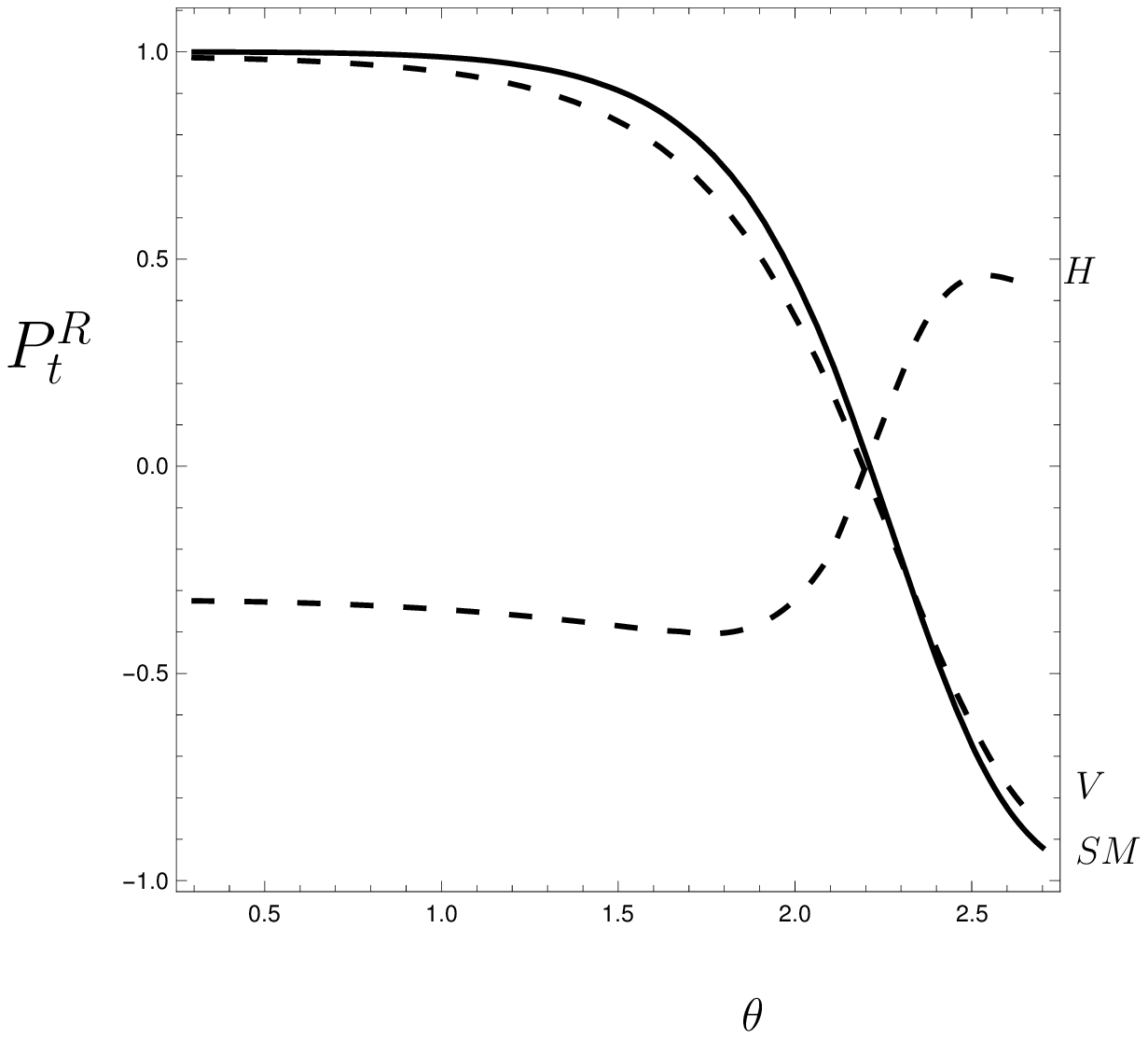 , height=20.cm}
\]\\
\vspace{-10cm}
\caption[1] {Energy and angular (at $\sqrt{s}=5$ TeV) dependences of the top quark polarization
in the $e^{-}_R$ polarized case,
in SM,  (H) with scalar Higgs type dark matter and (V) with vector dark matter.}
\end{figure}
\clearpage

\begin{figure}[p]
\vspace{-0cm}
\[
\hspace{-2cm}\epsfig{file=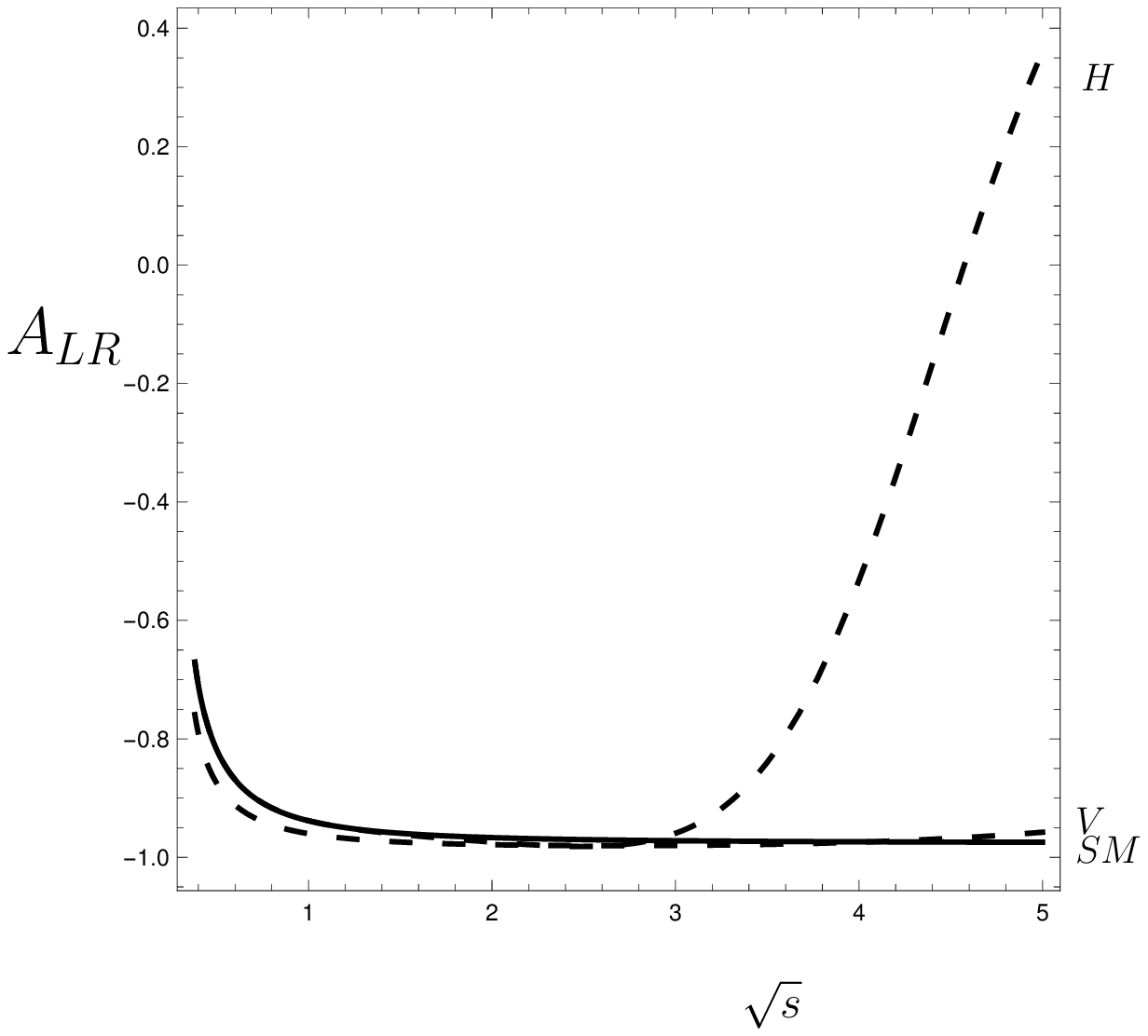 , height=20.cm}
\]\\
\vspace{-13cm}
\[
\hspace{-2cm}\epsfig{file=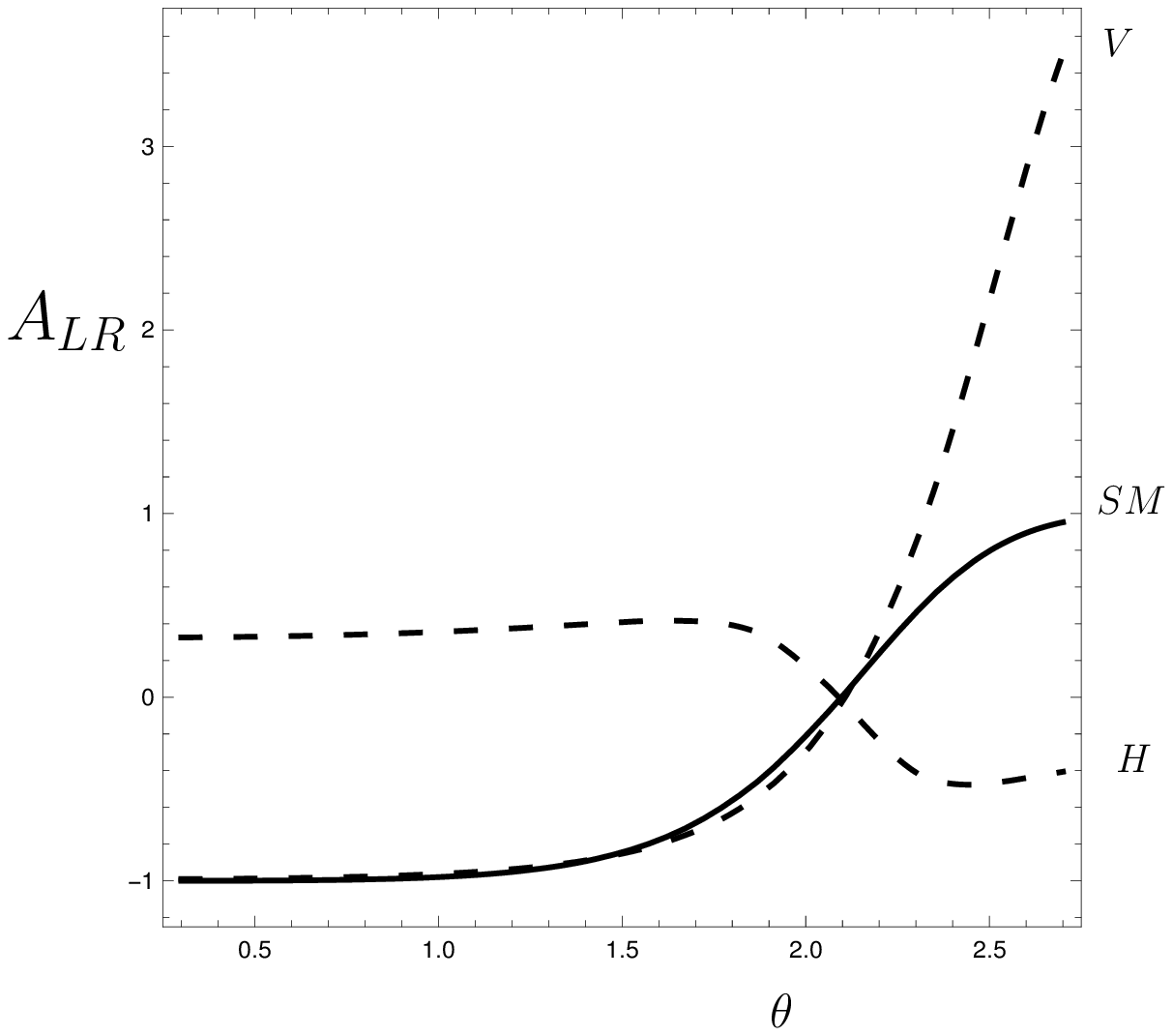 , height=20.cm}
\]\\
\vspace{-10cm}
\caption[1] {Energy and angular (at $\sqrt{s}=5$ TeV) dependences of the $e^{-}_L$-$e^{-}_R$
asymmetry of the top quark polarization, in SM,  (H) with scalar Higgs type dark matter and (V) with vector dark matter.}
\end{figure}
\clearpage

\begin{figure}[p]
\vspace{-6cm}
\[
\hspace{-2cm}\epsfig{file=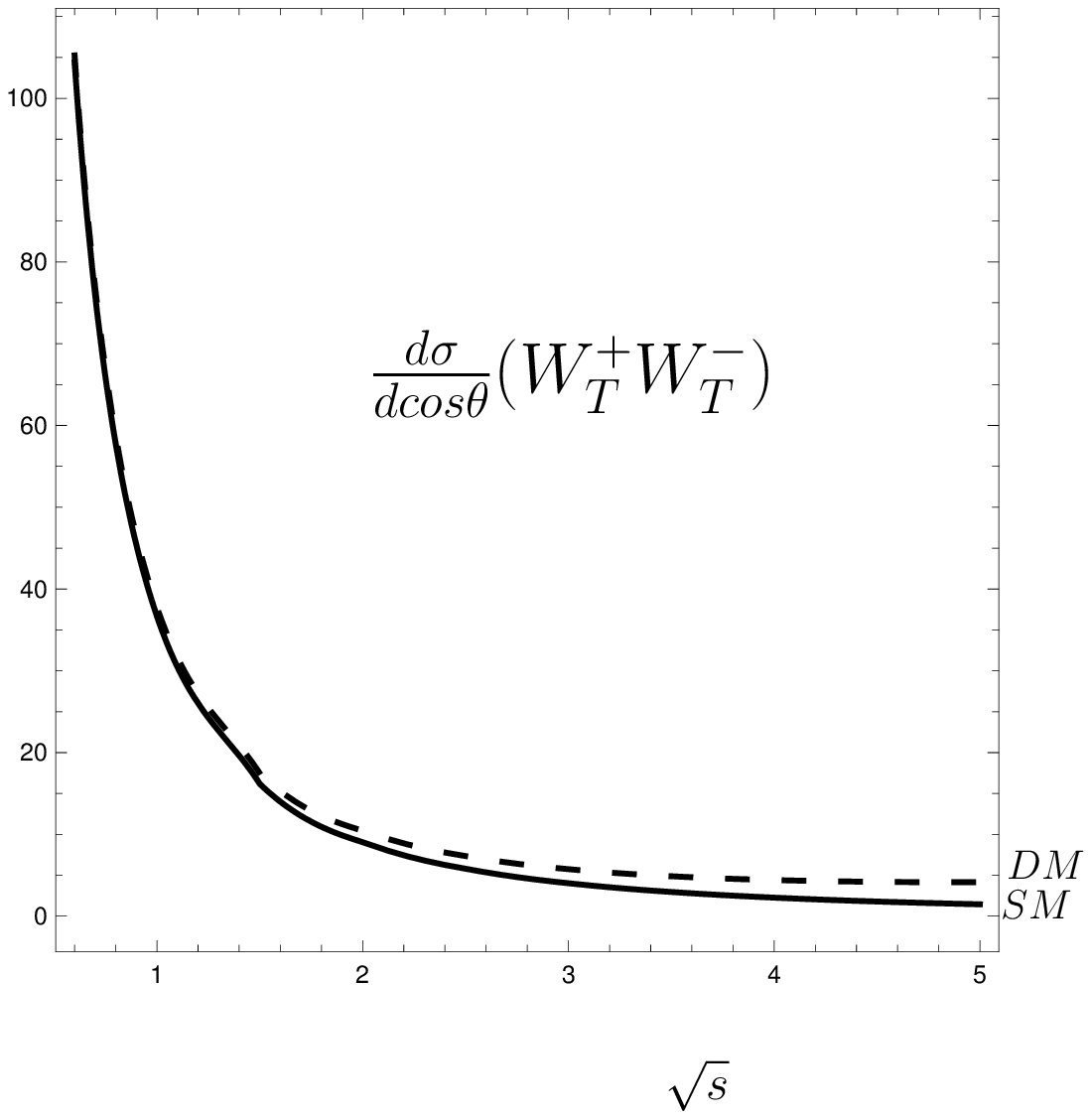 , height=18.cm}
\hspace{-4cm}
\epsfig{file=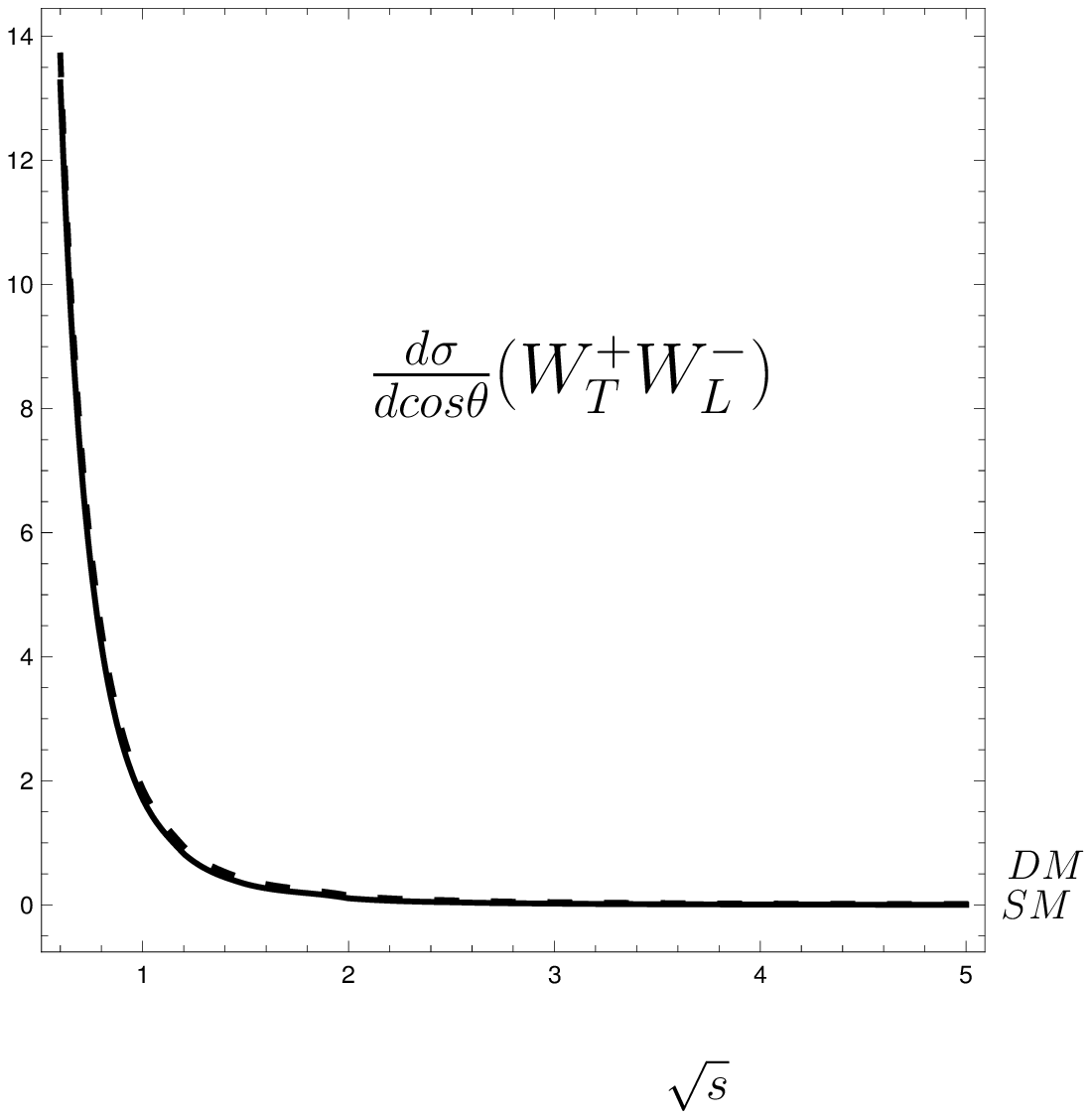 , height=18.cm}
\]
\\
\vspace{-12cm}
\[
\hspace{-2cm}\epsfig{file=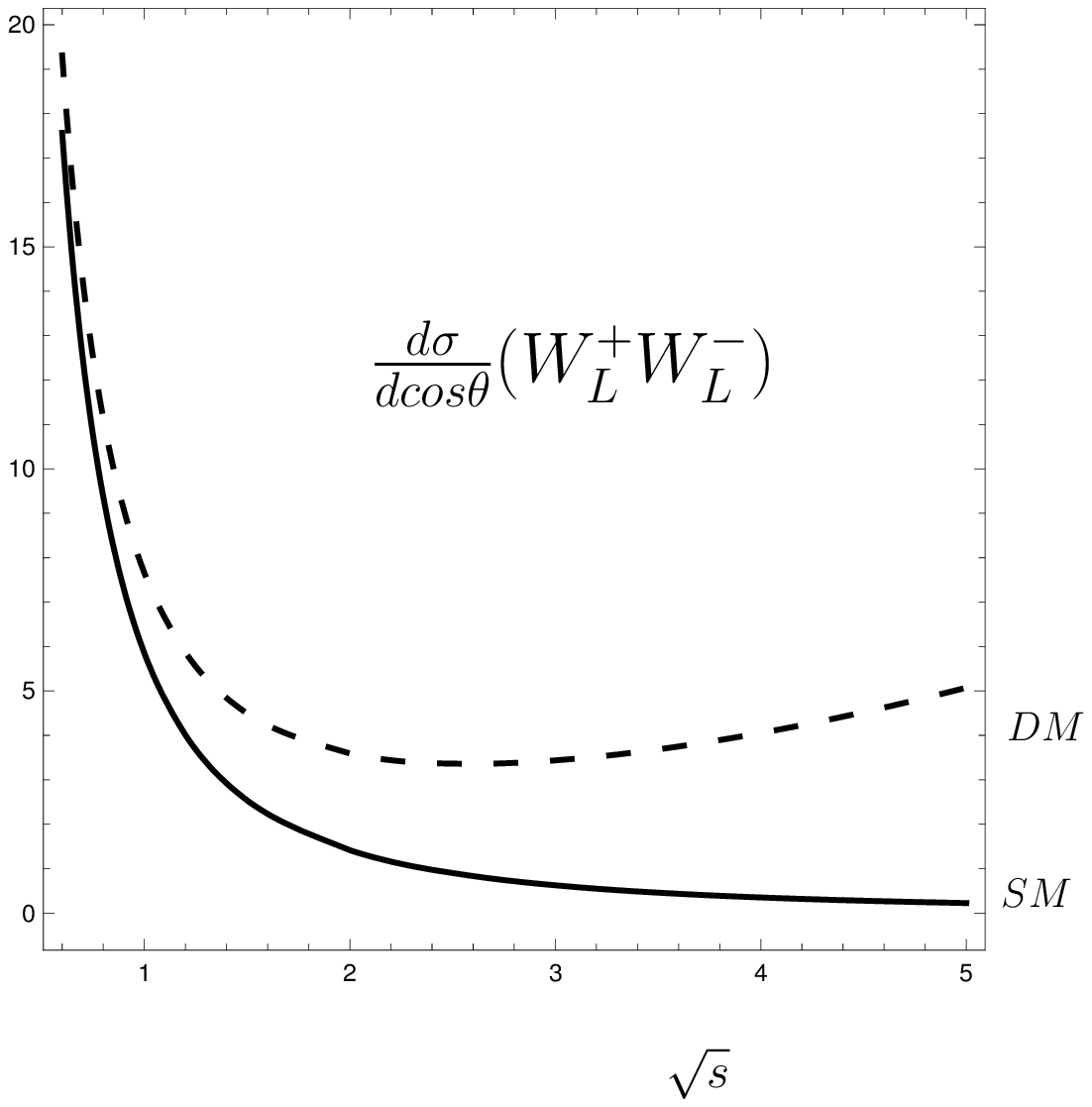 , height=18.cm}
\hspace{-4cm}
\epsfig{file=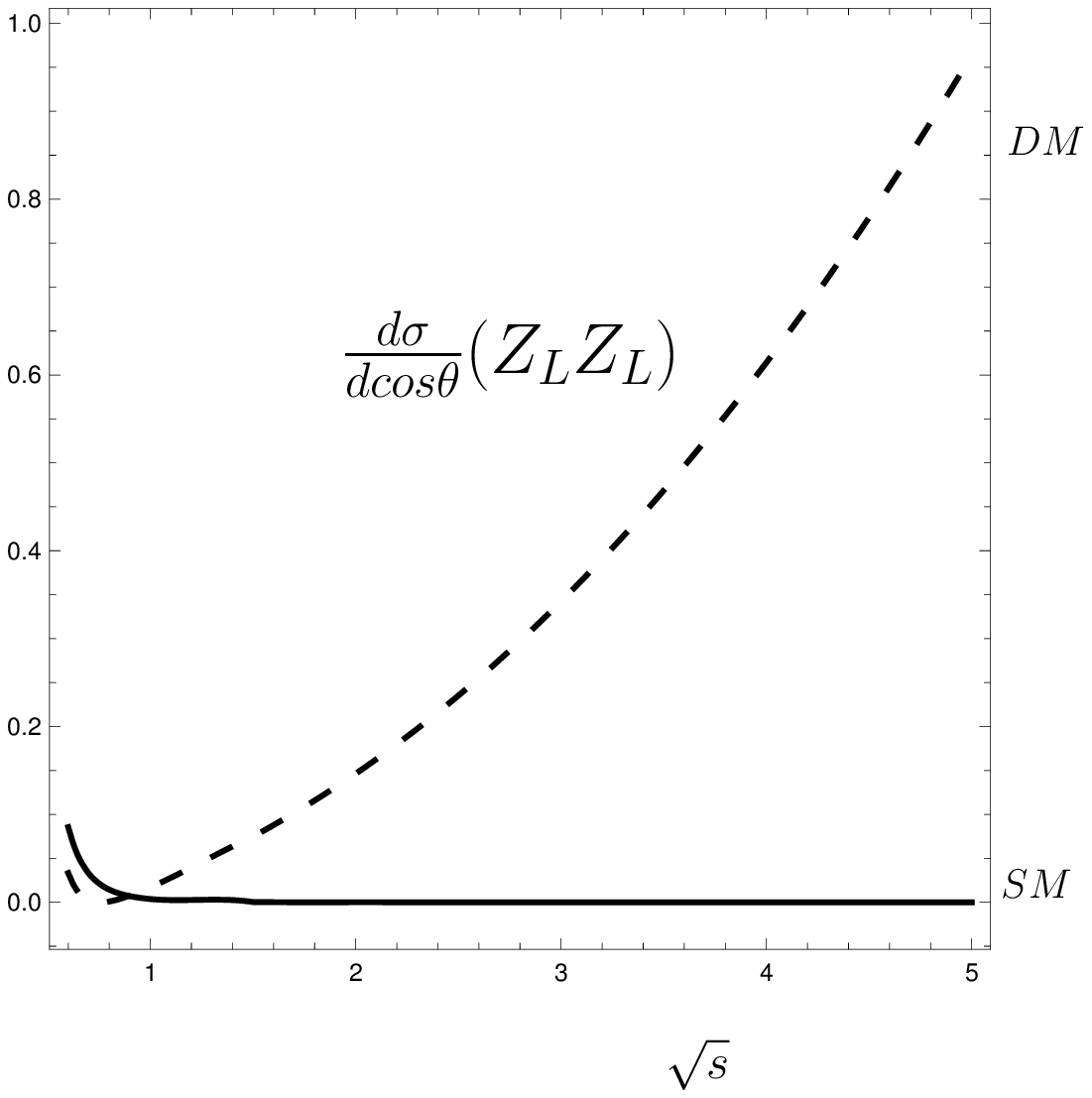 , height=18.cm}
\]\\
\vspace{-9cm}
\caption[1] {Energy dependence of the TT, TL, LL $e^+e^-\to W^+W^-$
and $e^+e^-\to ZZ$ cross sections.}
\end{figure}
\clearpage

\begin{figure}[p]
\vspace{-7cm}
\[
\hspace{-2cm}\epsfig{file=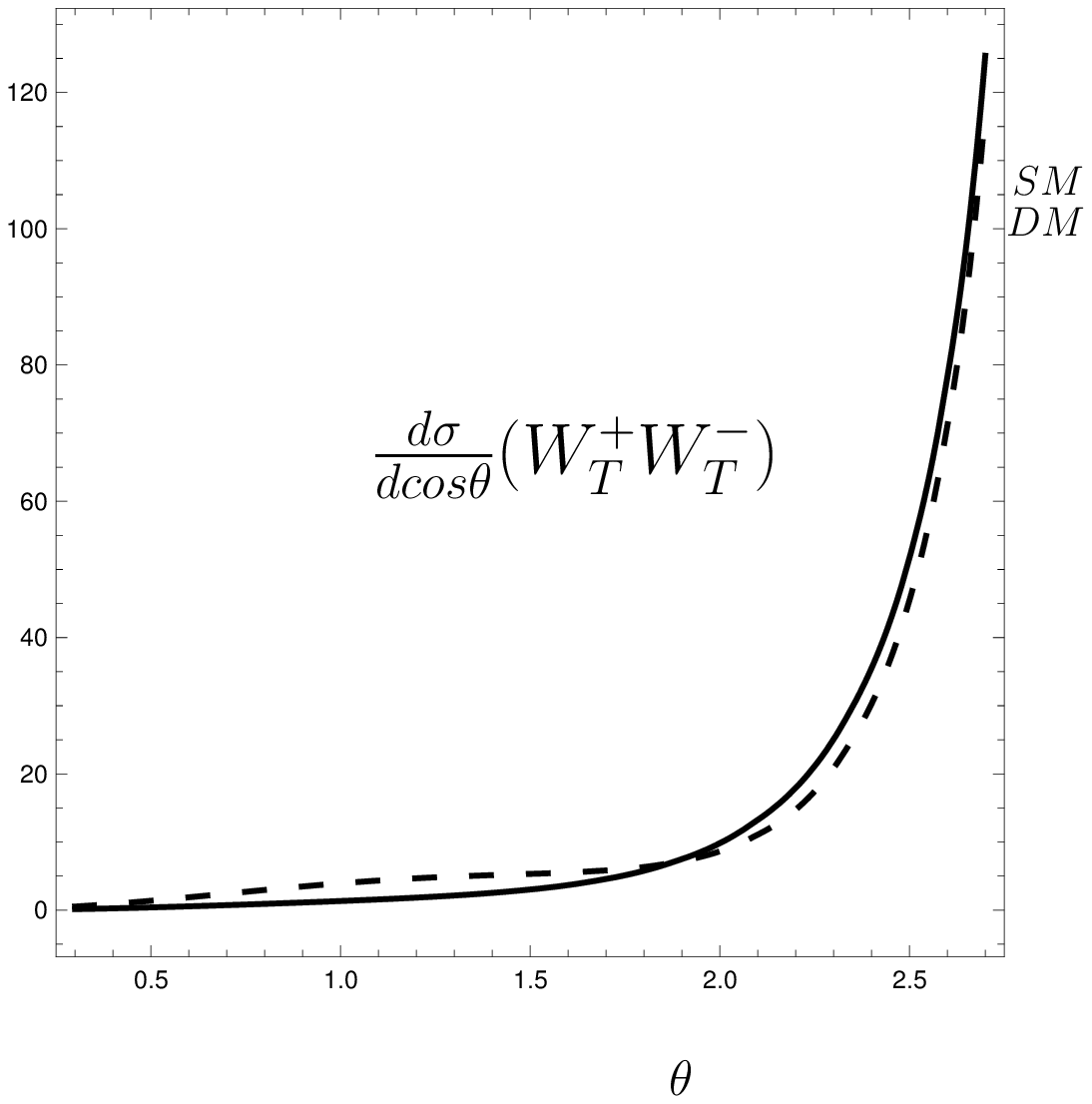 , height=18.cm}
\hspace{-4cm}
\epsfig{file=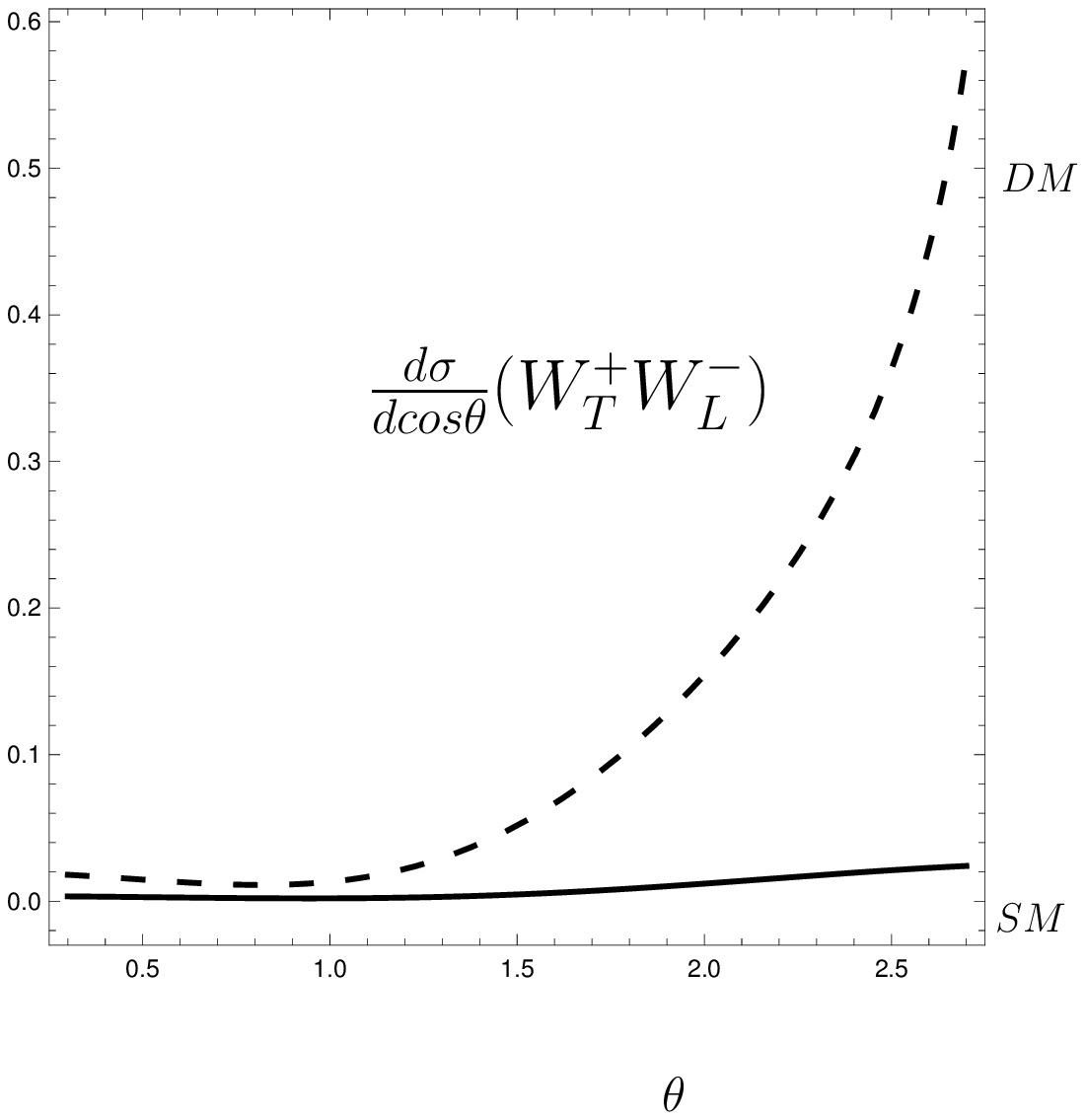 , height=18.cm}
\]
\\
\vspace{-12cm}
\[
\hspace{-2cm}\epsfig{file=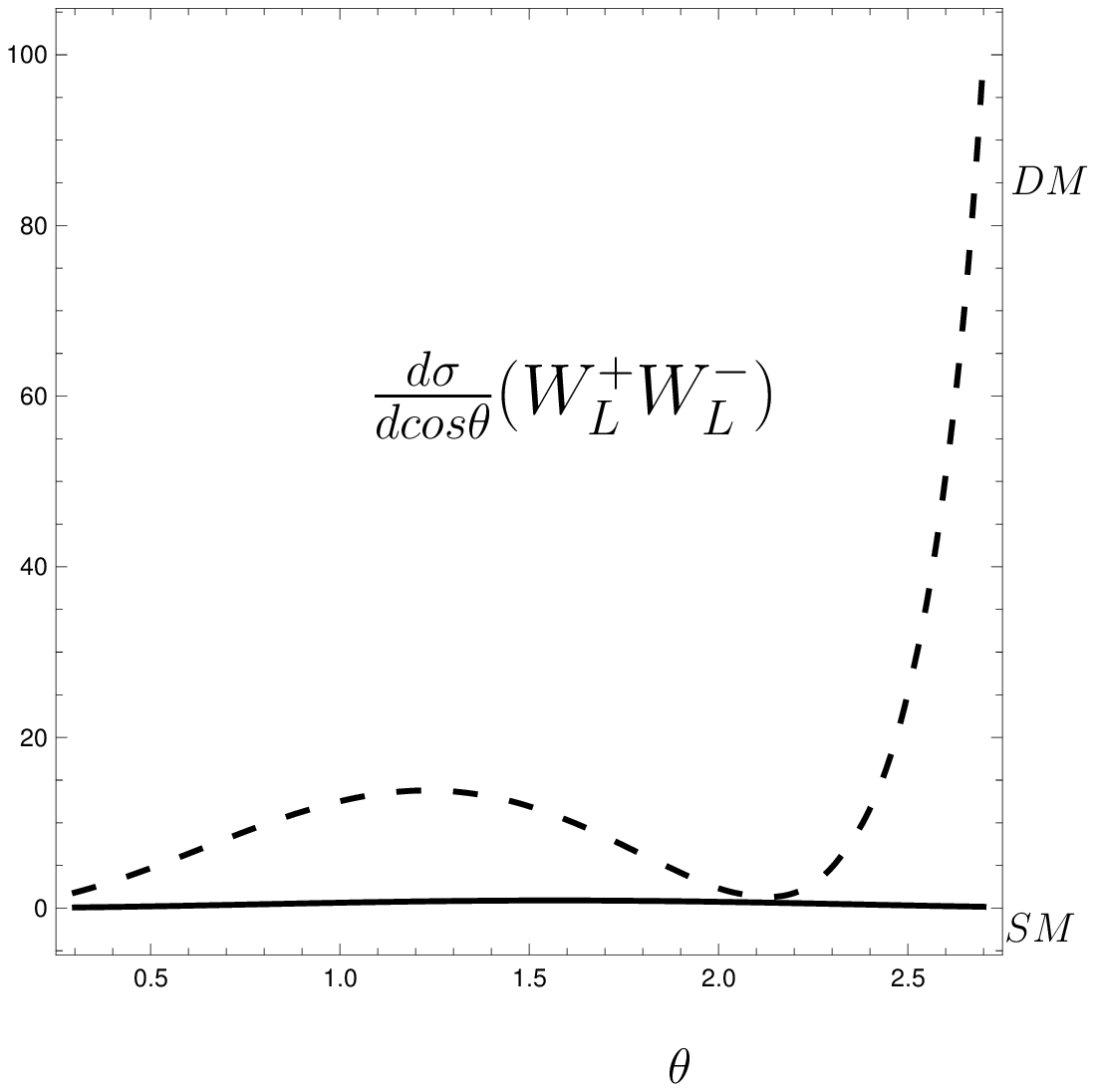 , height=18.cm}
\hspace{-4cm}
\epsfig{file=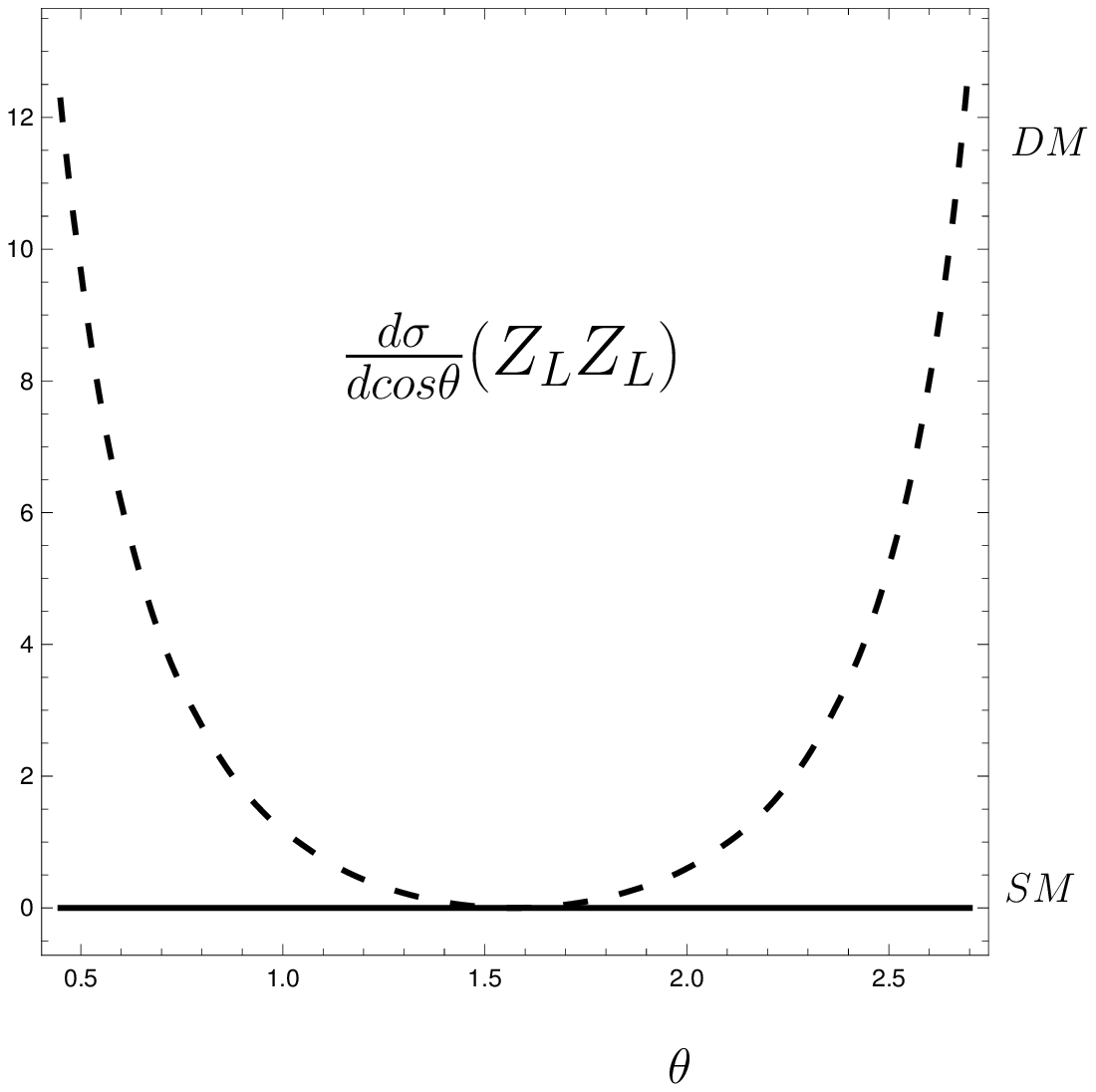 , height=18.cm}
\]\\
\vspace{-9cm}
\caption[1] {Angular dependence at $\sqrt{s}=5$ TeV of the TT, TL, LL $e^+e^-\to W^+W^-$
and $e^+e^-\to ZZ$ cross sections.}
\end{figure}
\clearpage

\end{document}